\documentclass[journal,onecolumn]{IEEEtran}
\usepackage{cite}
\usepackage{amsmath,amssymb,amsfonts}
\usepackage{graphicx}
\usepackage{textcomp}

\usepackage[usenames,dvipsnames,table]{xcolor}
\usepackage{float}
\usepackage{multirow}
\usepackage{setspace}
\usepackage{makecell}
\usepackage{algorithm}
\usepackage{algorithmicx}
\usepackage[noend]{algpseudocode}
\usepackage{esvect}

\makeatletter
\renewcommand{\Function}[2]{%
  \csname ALG@cmd@\ALG@L @Function\endcsname{#1}{#2}%
  \def\jayden@currentfunction{#1}%
}
\newcommand{\funclabel}[1]{%
  \@bsphack
  \protected@write\@auxout{}{%
    \string\newlabel{#1}{{\jayden@currentfunction}{\thepage}}%
  }%
  \@esphack
}

\definecolor{pastelG}{RGB}{154,255,153}
\definecolor{pastelR}{RGB}{255,204,201}

\def\BibTeX{{\rm B\kern-.05em{\sc i\kern-.025em b}\kern-.08em
    T\kern-.1667em\lower.7ex\hbox{E}\kern-.125emX}}

\graphicspath{{./figs/} {./figs/min/} {./figs/hp/} {./figs/flows/} {./figs/hp/ranking/} {./figs/hp/dist/} {./figs/hp/efficiency/}} 
\DeclareGraphicsExtensions{.pdf, .png, .svg}

\newcommand{\longcaption}[2]{\caption[#1]{\centering #1 \par \medskip #2}}

\begin{document}

\author{Luke McHale, 
        Paul V Gratz, and~Alex~Sprintson.
\thanks{The authors are with the Department
of Electrical and Computer Engineering, Texas A\&M University, College Station, Texas.}
}
 \title{Flow Correlator: A Flow Table Cache Management Strategy}


\maketitle
\begin{abstract}
    

Switching, routing, and security functions are the backbone of packet processing  networks. Fast and efficient processing of packets requires maintaining the state of a large number of transient  network connections. 
In particular, modern stateful firewalls, security monitoring devices, and software-defined networking (SDN) programmable dataplanes require maintaining stateful flow tables. These flow tables often grow much larger than can be expected to fit within on-chip memory, requiring a \emph{managed caching layer} to maintain performance. 

This paper focuses on improving the efficiency of caching, an important architectural component of the packet processing data planes. 
%
%
We present a novel predictive approach to network flow table cache management.  
Our approach leverages a Hashed Perceptron binary classifier as well as an iterative approach to feature selection and ranking to improve the reliability and performance of the data plane caches.

We validate the efficiency of the proposed techniques through extensive experimentation using real-world data sets. Our numerical results demonstrate that our techniques improve the reliability and performance of flow-centric packet processing architectures.

\end{abstract}




\section{Introduction}
Caching is an important component of many network devices.
Switching and routing functions are the backbone of all networks; however, an increasing amount of network functionality requires maintaining a flow state for transiting connections.
Modern stateful firewalls, security monitoring devices, and software-defined networking require maintaining stateful flow tables.
These flow tables often grow much larger than can be expected to fit within on-chip memory, requiring a managed caching layer to maintain performance.
This caching layer is often obfuscated and embedded between the control and data plane layers.

The design space of network devices is continually evolving with trade-offs in performance and reconfigurability.
A common architectural advantage of network processors is to provide hardware-assisted table lookup extensions for both on-chip and memory-backed tables.  
Network functions requiring large flow tables, often consisting of millions of entries, require managing both on-chip and off-chip table resources.
To meet performance and cost requirements, caching mechanisms are commonly employed.
Caching a flow table using on-chip table resources preserves memory bandwidth and reduces latency, but introduces the complexity of effective cache management.





In processor microarchitecture, it has long been known that software inherently exhibits locality that can be leveraged at run-time.
Cache management for network flow tables has historically seen less research interest than in processor microarchitecture.
Insight into state-of-the-art flow table management tends to be particularly shrouded in proprietary implementations.
Existing flow table cache management heuristics are primarily based on the Least Recently Used (LRU) replacement as well as explicit protocol-based bypass mechanisms.




This paper aims to explore if locality in network traffic can be leveraged with more advanced cache management techniques.
We review several modern cache management approaches in microarchitecture, highlighting our progress in exploring flow table cache management.
The contributions of this paper include:

\begin{enumerate}
\item Flow table cache management limit study using Belady's MIN replacement algorithm on CAIDA network exchange traffic.
\item First work to apply a Hashed Perceptron binary classifier to network flow table cache management.
\item Iterative approach to feature selection and ranking in the context of Hashed Perceptron binary classifiers.
\item Discussion of feature roles, granting further perspective into the dynamics of a Hashed Perceptron binary classifier.
\end{enumerate}


\section{Motivation}

To bring perspective to the objective metrics used in this paper, Section \ref{sec:hp-hit-rate} first describes the significance of cache hit-rate to system performance.
In order to understand the upper bound on room for improvement to cache hit-rate, we start with an optimality study in Section \ref{sec:hp-opt}.
Cacheable packet inter-arrival patterns and their timescales are further motivated in Section \ref{sec:hp-patterns}.
Section \ref{sec:hp-stack} reviews well-known stack-based cache management algorithms.
Limitations to traditional approaches adapted to caching in networking are outlined in Section \ref{sec:hp-failures}.
Finally, we wrap up by introducing modern multi-perspective approaches to cache management using the hashed perceptron binary classifier in Section \ref{sec:hp-perceptron}.

\subsection{Significance of Cache Hit-Rate} \label{sec:hp-hit-rate}
System architects need to understand the impacts and potential drawbacks of caching, including estimations of processing latency and throughput.
Caching allows mitigating slow-path processing requirements by taking advantage of locality.
The actual performance benefit of a proposed caching implementation depends on the ability to capture the effective working set available in the workload.
It is important to quickly assess the potential expected performance improvement granted by a proposed caching architecture.

\begin{equation}
APPT \approx AMAT = T_{fast} + Miss Rate \times T_{slow}
\label{equ:f-estimate-performance}
\end{equation}

Equation \ref{equ:f-estimate-performance} rephrases average memory access time (AMAT) to approximate the sensitivity of cache hit rate on average packet processing time (APPT).
As accumulated packet processing effort roughly tracks processing latency, APPT can be approximated through a first-order processing latency estimation.
In the context of packet processing, $T_{slow}$ represents the average latency (or effort) exerted while traversing the slow path.
$T_{fast}$ represents the average latency (or effort) consumed when a cache hit enables fast-path processing.
It is expected that the processing effort between fast and slow paths could be an order or two apart -- further motivating the desire to maximize cache hit rate.

The wide variety of stateful network functions as well as packet processing architectures, make it difficult to settle on a specific known ratio between $T_{fast}$ and $T_{slow}$. 
For simplicity, this first-order sensitivity estimation is scoped to a single data plane packet processing slice.
A more detailed analysis approach may be needed to take into account the intricacies of parallel processing pipelines.


As is common in processor microarchitecture cache management, small percentile average hit-rate improvements often translate into notably significant ten-percentile performance gains.
While the sensitivity estimation is ultimately left to the designer, it is conceivable that single-digit hit-rate improvements presented in this paper may translate to a multiple-digit performance speedup for stateful network functions reliant on caching.

\subsection{Cache Optimality Study} \label{sec:hp-opt}
Originally developed to study optimal page replacement for virtual memory systems, Belady's Minimal page replacement algorithm (MIN) has been applied to cache replacement in processor microarchitecture \cite{belady-min, belady-replacement, mattison-stack}.
Belady's MIN algorithm effectively orders the replacement stack by the next furthest access, maximizing the cache hit rate.

Outside of a few specialized domains with bounded working sets or known access patterns, an optimal replacement ranking is usually impractical to build in hardware.
Unpredictable random events will be capturable by MIN, but elusive to practical implementations which rely on pattern history.
As an upper bound, Belady's MIN provides useful insight to cache replacement as well as potential headroom to improve cache hit rate.

\begin{figure}[ht]
\centering
\includegraphics[width=\linewidth]{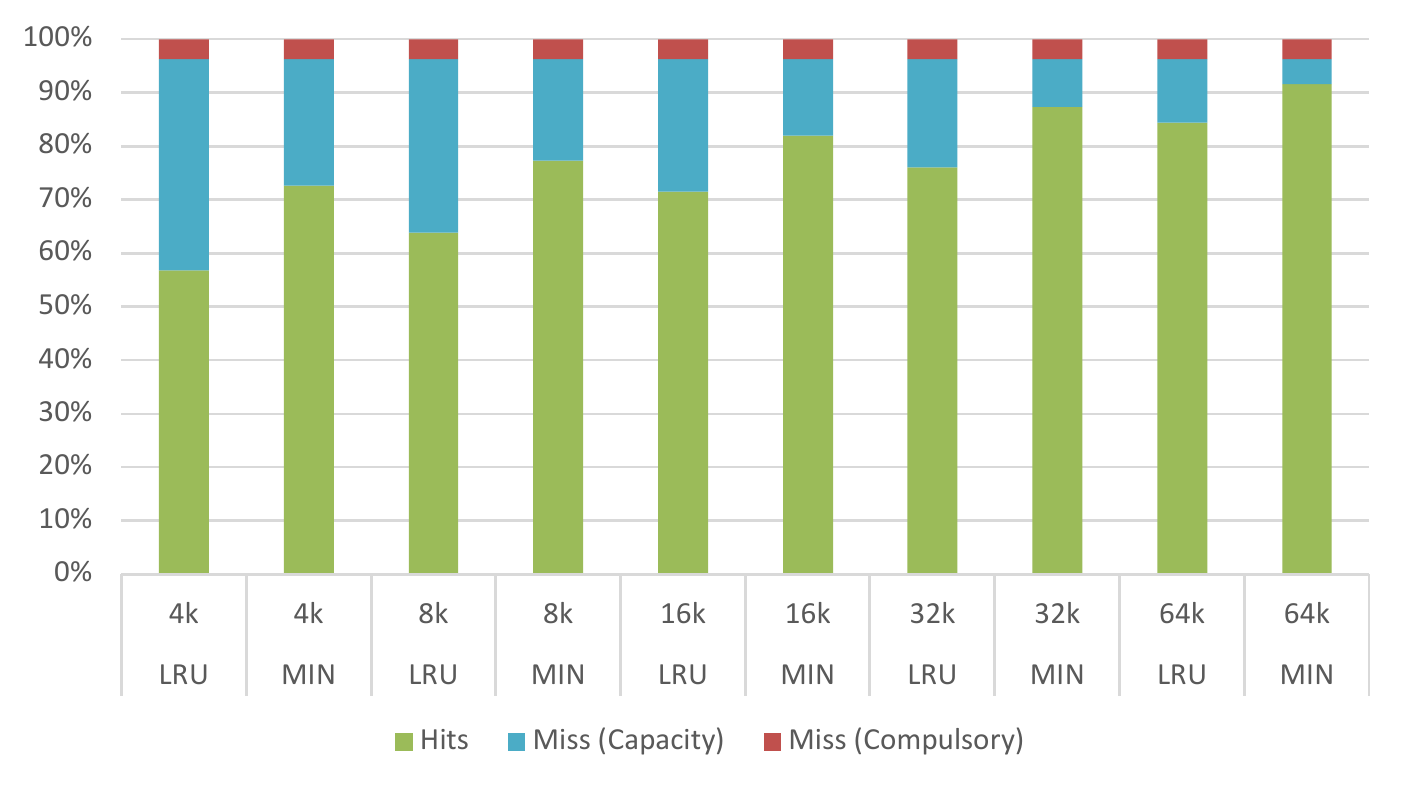}
\longcaption{Belady's MIN Optimal Replacement}{
Fully associative flow table cache hit-rate simulation.  CAIDA Equinix Sanjose January 2012 dataset.}
\label{fig:hp-belady}
\end{figure}

Figure \ref{fig:hp-belady} shows the cache hit-rate for both oracle Belady's MIN algorithm and a baseline fully-associative LRU implementation.
While compulsory misses (misses caused by new flows) are unavoidable, MIN's capacity misses are entirely limited by the cache capacity and not also the replacement algorithm.
The hit-rate delta between MIN and LRU is an upper bound to the best possible hit-rate improvement achievable by improving the cache management algorithms.

At 4k flow entries, LRU achieved $15.8\%$ fewer hits relative to MIN, with $23.7\%$ true capacity misses remaining.
This gap reduced to $7.2\%$ fewer hits at 64k entries, with only $4.7\%$ capacity misses remaining.
It is expected that there will always be a gap between realistic cache management algorithms and oracle algorithms.
However, the following sections aim to investigate the potential for improvement.

This limit study suggests there might be a potentially cacheable working-set \cite{denning-working-set} larger than what LRU is able to capture.
However, there is no guarantee that the gap between MIN and LRU is practically achievable.
Intuitively, MIN also confirms greater potential improvement as cache size decreases.
Section \ref{sec:hp-performance} further discusses the significance of hit rate including the expected impact on overall system performance.


\subsection{Flow Patterns} \label{sec:hp-patterns}
While network traffic, particularly transport layer protocol behavior, is relatively well understood and modeled \cite{markov-traffic-modeling};
we theorize that network flows also exhibit locality driven by program phases on connection endpoints.
Examples of side-channel attacks \cite{ssh-sidechannel} further support inherent end-host program behavior embedded within packet inter-arrival patterns.

\begin{figure*}[t]
\centering
\begin{tabular}{cc}
\small
\includegraphics[width=87mm]{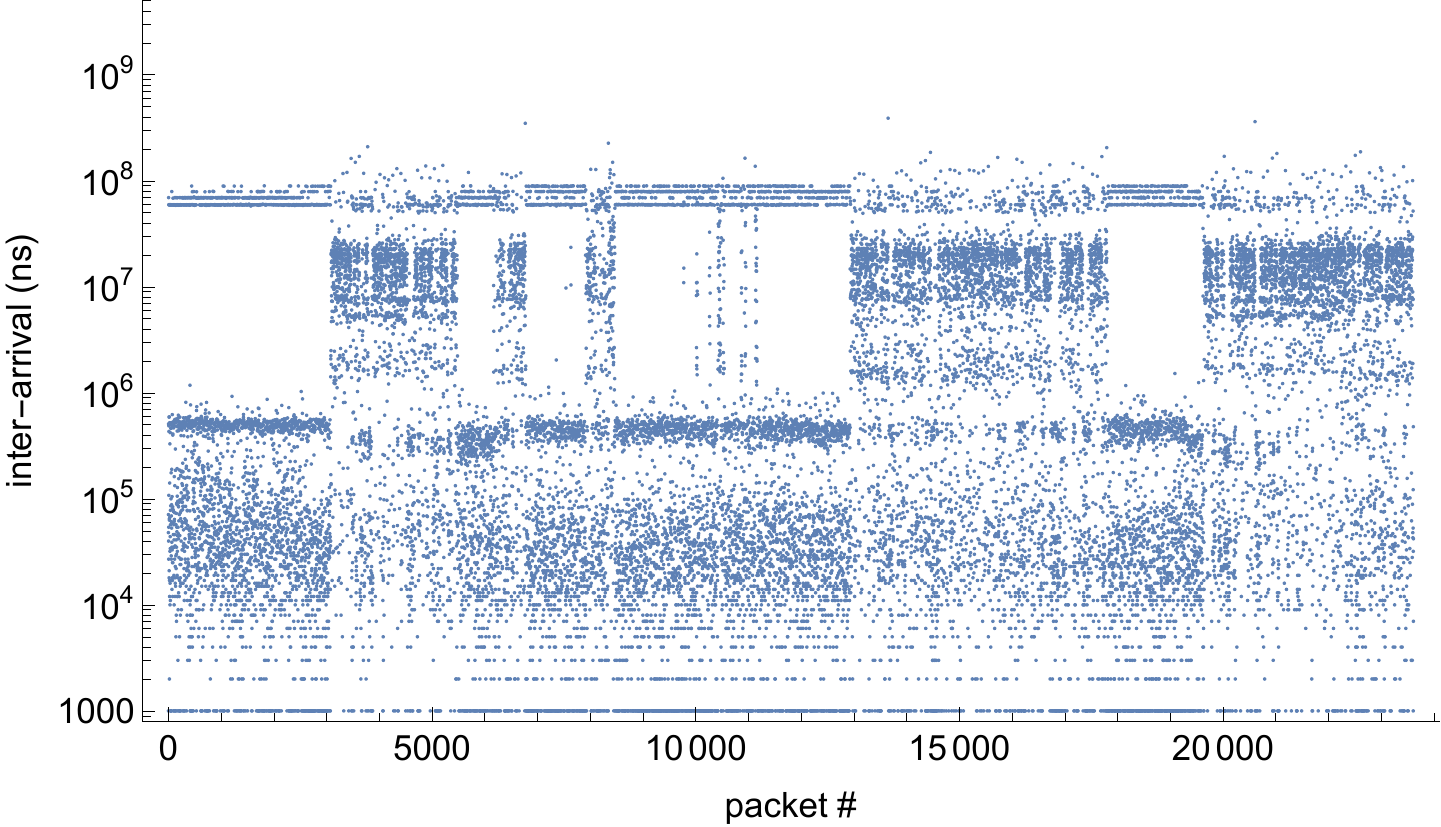} & \includegraphics[width=87mm]{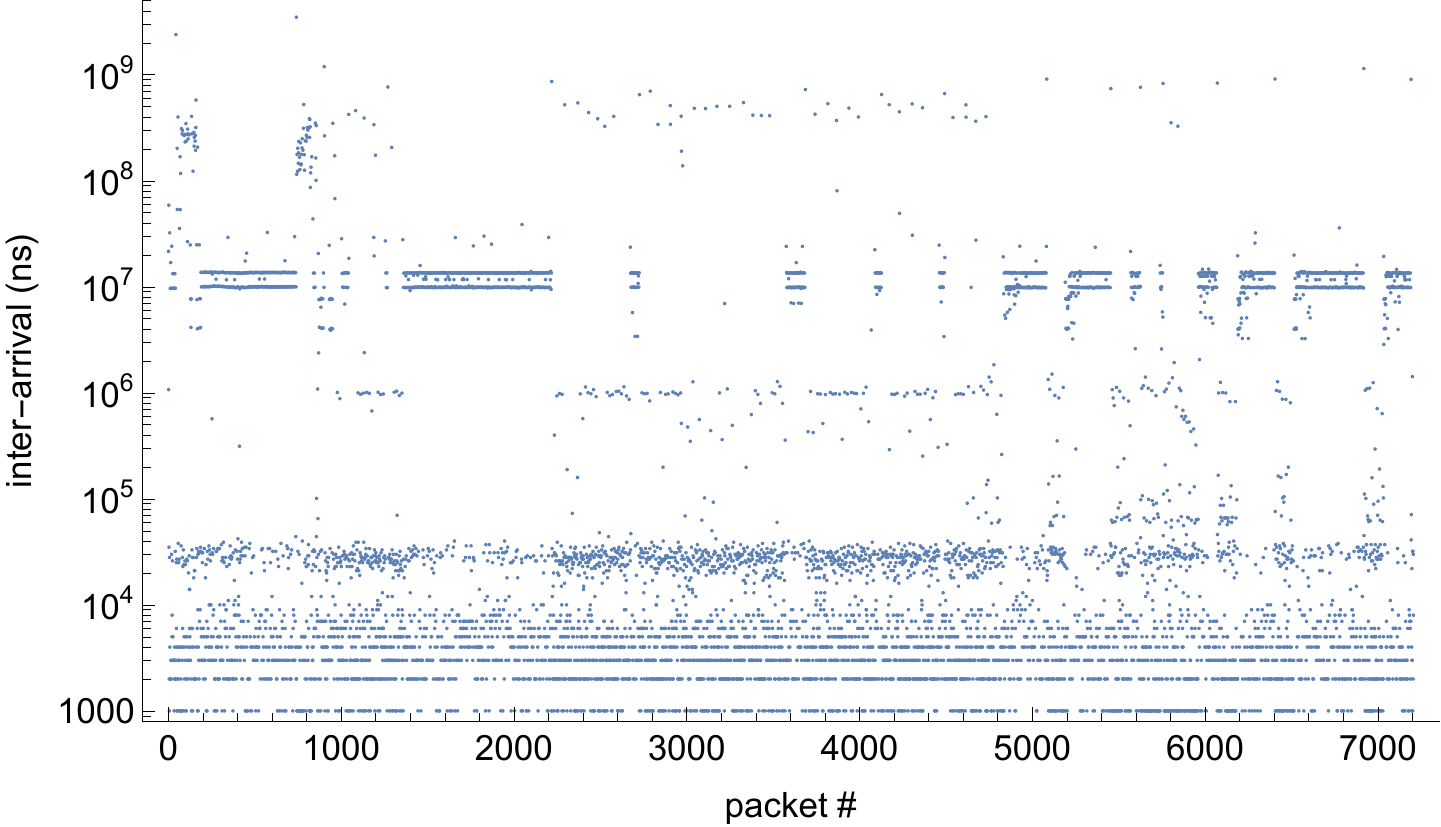} \\
(a) & (b) \\[6pt]
\includegraphics[width=87mm]{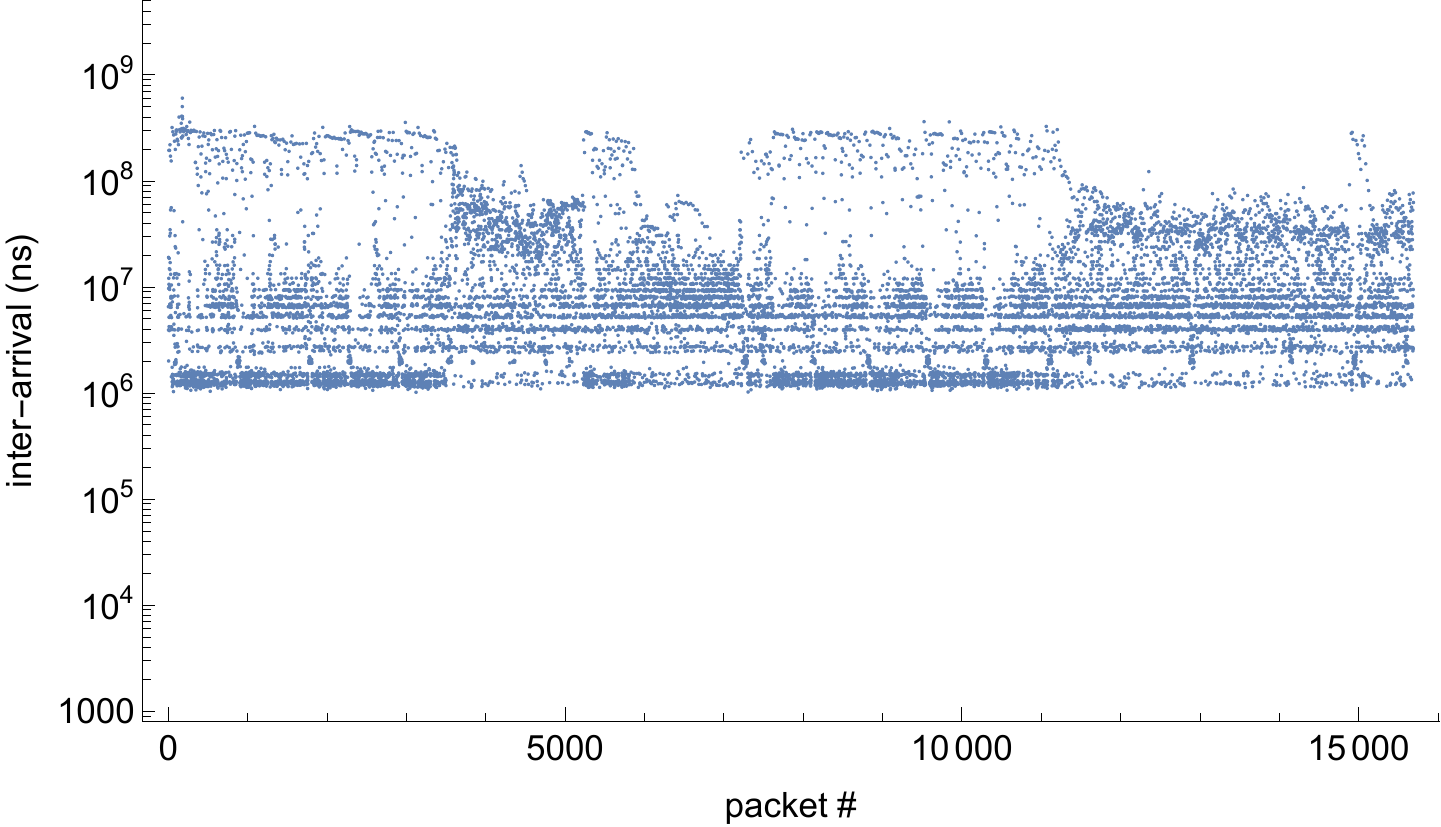} & \includegraphics[width=87mm]{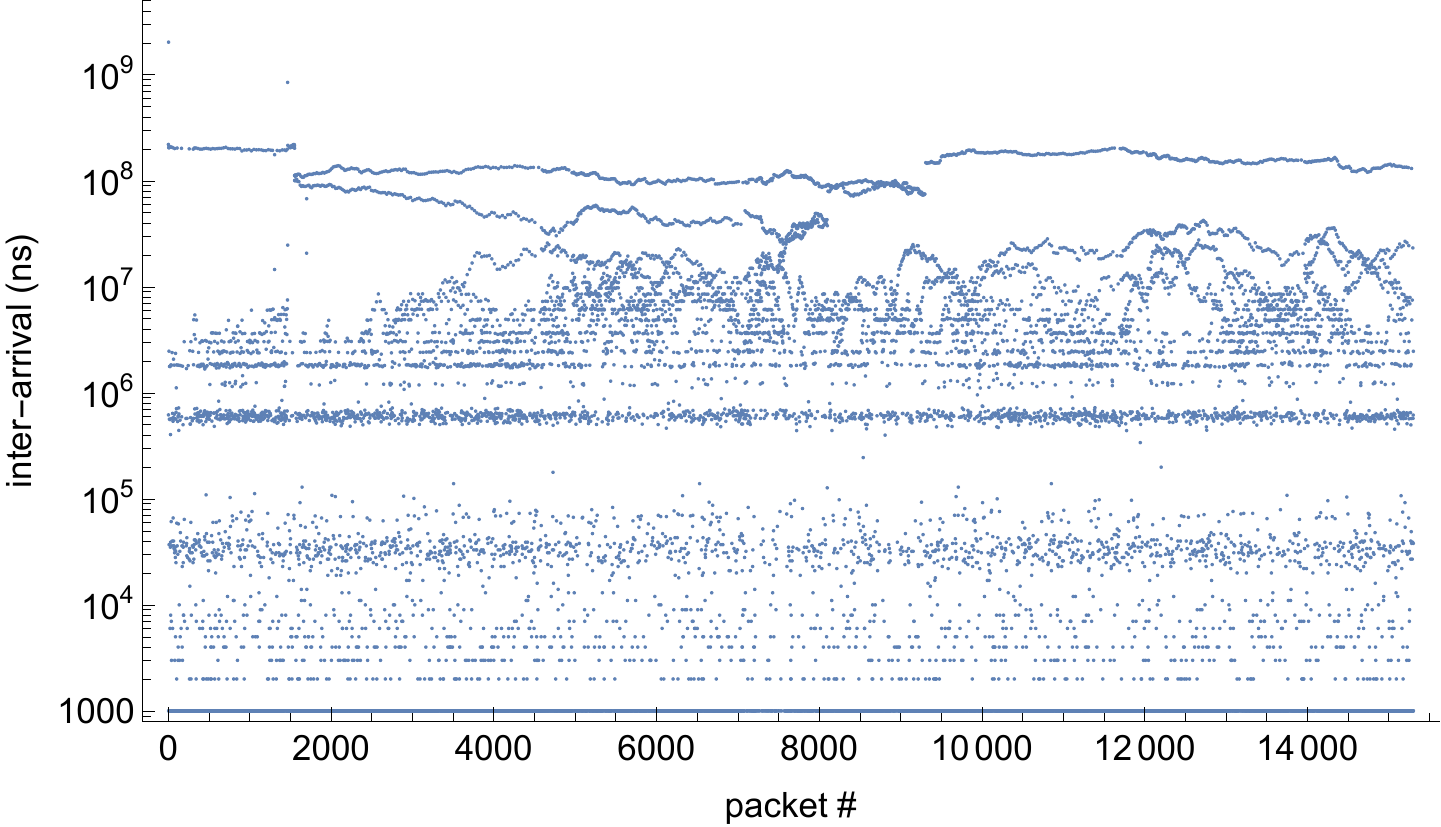} \\
(c) & (d) \\[6pt]
\end{tabular}
\longcaption{Interesting Packet Inter-arrival Patterns by Flow}{
The log-scale y-axis shows inter-arrival delay between packets (in nanoseconds), while the x-axis is packets since the start of the flow.}
\label{fig:hp-flows}
\end{figure*}

Figure \ref{fig:hp-flows} showcases four hand-selected flows extracted from the 2012 Sanjose CAIDA dataset\footnote{CAIDA Equinix Sanjose January 2012 PCAP trace: equinix-sanjose.20120119.}
in order to showcase a few representative multi-modal patterns.
These plots should be interpreted not strictly as a linear time-series, but similar to a waterfall plot drawn across the x-axis, where the y-axis is a probability density of delays between packets.

Clusters of packet inter-arrival below the approximate transport layer Round Trip Time (RTT)\footnote{RTT is a measure of round-trip latency between two end-hosts -- commonly on the order of 10ms to 100ms for Internet traffic.}
can be considered packet bursts.
Similarly, inter-arrival delays around RTT can be roughly attributed to short-term transport-layer behavior.
Finally, inter-arrival delays far exceeding RTT processing are likely attributable strictly to application level behavior, with an upper bound on flow keep-alive\footnote{Keep-Alive is a common transport-layer timeout concept, primarily used to ensure a dormant flow remains tracked in stateful Flow Tables -- usually on the order of seconds.}
delays.

While packet inter-arrivals are inherently noisy, these flows appear to exhibit potential predictable, modal, almost cyclic patterns.
We theorize that the large discrete shifts in packet inter-arrival delays are driven by end-host application behavior and are not simply transients in network conditions.
In the context of cache management, these shifts in application behavior most notably impact packet burst duration and inter-frequency.
Here we aim to explore whether there are cacheable patterns above what the canonical LRU replacement policy is able to identify.


\subsection{Stack-based Algorithms} \label{sec:hp-stack}
Cache management algorithms that maintain ranked order for replacement or insertion are formally referred to as stack-based algorithms.
The Least Recently Used (LRU) algorithm intuitively maintains a replacement stack ordered by the last access time.
Specifically, LRU is analogous to Belady's MIN reversed in time -- ranking past instead of future accesses.
Ultimately, LRU is effective when past history leads to a good prediction of the future.

LRU assumes reuse is always likely, inserting all new items in the most recent position.
Adaptive insertion policies such as Dual Insertion Policy (DIP) \cite{patt-joel-dip}, allow insertion into either the top or bottom of the LRU cc stack.
DIP is similar to bypassing the cache entirely, but does grant the demoted insertion a chance to become promoted.
Adjusting the promotion/demotion logic of LRU can improve cache efficiency in the context of the unequal probability of reuse \cite{jiminez-lru-promotion}.

Replacement policies based on reference counts have proven useful as a means to capture temporal patterns.
Cache Bursts \cite{doug-cachebursts} introduced both RefCount and BurstCount as mechanisms to estimate reuse predictions from cache entry hit counters.
Reference interval counting mechanisms continue to be researched and improved, leading to Re-Reference Interval Prediction (RRIP) \cite{joel-rrip} and Signature-based Hit Predictor (SHiP) \cite{joel-ship}.

Branch prediction research has long been combining multiple information sources, leading to mechanisms to combine or prioritize multiple predictors.
The TAGE predictor is similarly able to combine multiple predictions in a ranking mechanism to select the most likely accurate prediction \cite{seznec-tage2}.
While the Hashed Perceptron mechanism has been steadily gaining momentum as a robust means to consider multiple potentially indicative features together as a whole \cite{jiminez-hp-bp, tarjan-hp-gshare}.

The Hashed Perceptron technique originated in branch prediction but has been successfully applied to cache management \cite{samira-hp-sampling, teran-reuse-prediction,  chirp-reuse, teran-multiperspective}.
Recent research has also leveraged the Hashed Perceptron structure as a filtering mechanism to improve prefetching quality \cite{eshan-hp-prefetch}.
Similar efforts towards prefetching filtering were also accomplished using Bloom Filters \cite{i-spy}.

Several research efforts have applied offline deep learning techniques towards cache replacement \cite{lin-deep-replacement}.
Similarly, Hawkeye \cite{lin-belady-hawkeye} leveraging a time-delayed Belady's min algorithm as an online replacement prediction feedback mechanism.

\subsubsection{Preliminary Exploration} \label{sec:hp-failures}
In processor microarchitecture, access patterns are constantly shifting, depending on program behavior.
Similarly in networking, flows also exhibit modes of operation depending on the protocol transport layer as well as end-host program phases.
In the context of a flow table cache replacement policy, LRU does a decent job of capturing short-term packet bursts.
However, LRU's ranking can also be polluted by dormant, low-bandwidth flows competing for cache resources.
LRU has no way to differentiate reuse probability, treating all flows as equally likely.

The insertion policy is especially critical to flow table cache management due to the short lifetime of most flows.
Most flows are short-lived with the majority of bandwidth attributed to a small subset of flows \cite{mchale-bloom}.
However, it is particularly challenging to predict which flows will be short-lived early in the flow life-cycle.

Inspired by prior works in CacheBursts \cite{doug-cachebursts} and SHiP \cite{joel-ship}, cache entry hit counters seemed well-suited as a means to track packet bursts.
However, early attempts at leveraging CacheBursts and SHiP directly as cache management strategies resulted in performance degradation compared to baseline LRU implementations when applied to flow table cache management.
That said, CacheBursts ended up being a notably informative feature component in our proposed final design.

Our single-perspective cache management approaches were not able to outperform LRU replacement when applied to managing flow tables.
Packet arrival variability inherent in networking adds significant noise to simple pattern predictors such that simple predictor tables alone could not outperform a baseline LRU implementation.
It became evident that a more robust pattern correlation mechanism is needed to rank reuse probability amongst noisy inter-arrival patterns.


\subsection{Hashed Perceptron Binary Classifier} \label{sec:hp-perceptron}

The Hashed Perceptron binary classifier has been successfully used in processor microarchitecture as an approach to improve cache management heuristics.
While research on how to integrate and apply this prediction technique has extended over a decade, the technique is now being shipped in modern processors in recent years.

Much like modern machine learning approaches, the Hashed Perceptron improves overall reuse prediction quality by combining multiple predictive features into a singular weighted prediction.
While conceptually similar to perceptrons in a single-layer neural network, the Hashed Perceptron has the significant advantage of being efficient to implement in both hardware and software.
The Hashed Perceptron is also paired with an adaptive training technique to adjust to changing run-time behavior.

Differing from a traditional Perceptron model, which scale analog values, requiring integral meaning for each perceptron input.
The Hashed Perceptron is able to map disjoint binary inputs to a corresponding correlation using the feature's table.
The correlation output can be interpreted as an integral value, expressing a historical correlation of a particular feature to the inference question at hand.
During inference, combining multiple features' independent correlations result in a strengthened overall prediction.

The processor microarchitecture community has demonstrated the Hashed Perceptron is well suited for reuse prediction, performing especially well in noisy multicore cache hierarchies.
This paper outlines our exploration into adapting the Hashed Perceptron mechanism to flow table cache management.


Our early exploration discovered that relying on any single feature is fragile, especially when applied to network flow table cache management.
Traditional single-feature approaches failed to surpass the consistent performance of LRU when applied to network caching.
The variety of traffic patterns encountered in network cache management demands a general approach to consistently approximate a good working set.

Caching will continue to be a fundamental component of network data planes as a means to amortize costly control path decisions.
The trend in networking to support programmability of generic network functions increases control path complexity.
Advances in cache management strategies leveraged in modern processor microarchitecture are absent in the networking domain.
Clearly, there is a gap between Belady and LRU, can we do better?

\section{Design} \label{sec:hp-design}
One of the goals of this work is to explore a generalized approach to flow table cache management.
Stateful flow tables are core components in the classification pipeline for firewalls and edge security devices.
However, caching can be applied generically to any large table supporting a classification pipeline.
While the focus of this paper is on improving cache management for stateful flow tables, the techniques we propose can be applied to assist any data plane table considered for caching.

\begin{figure}[ht]
\centering
\includegraphics[width=0.6\linewidth]{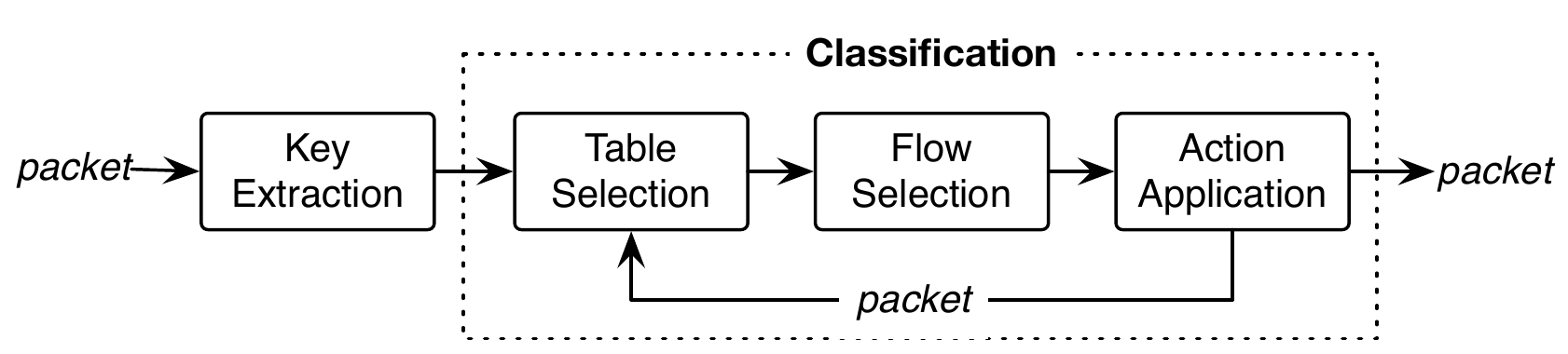} 
\caption{Data Plane Classification Processing}
\label{fig:hp-classification}
\end{figure}

Figure \ref{fig:hp-classification} portrays an abstract cycle of table selection, flow selection, and action application as the underlying fundamental classification abstraction.
Network functions are composed by chaining multiple logical rounds of classification across relevant flow-state tables.
Network functions that also manipulate packets often also queue data plane and packet actions\footnote{OpenFlow specifies certain actions apply immediately, while others may be queued in an action-set with notable complications.}.
Fundamentally, SDNs aim to expose table management and packet manipulation for a composable packet processing pipeline.



\begin{figure}[ht]
\centering
\includegraphics[width=0.8\linewidth]{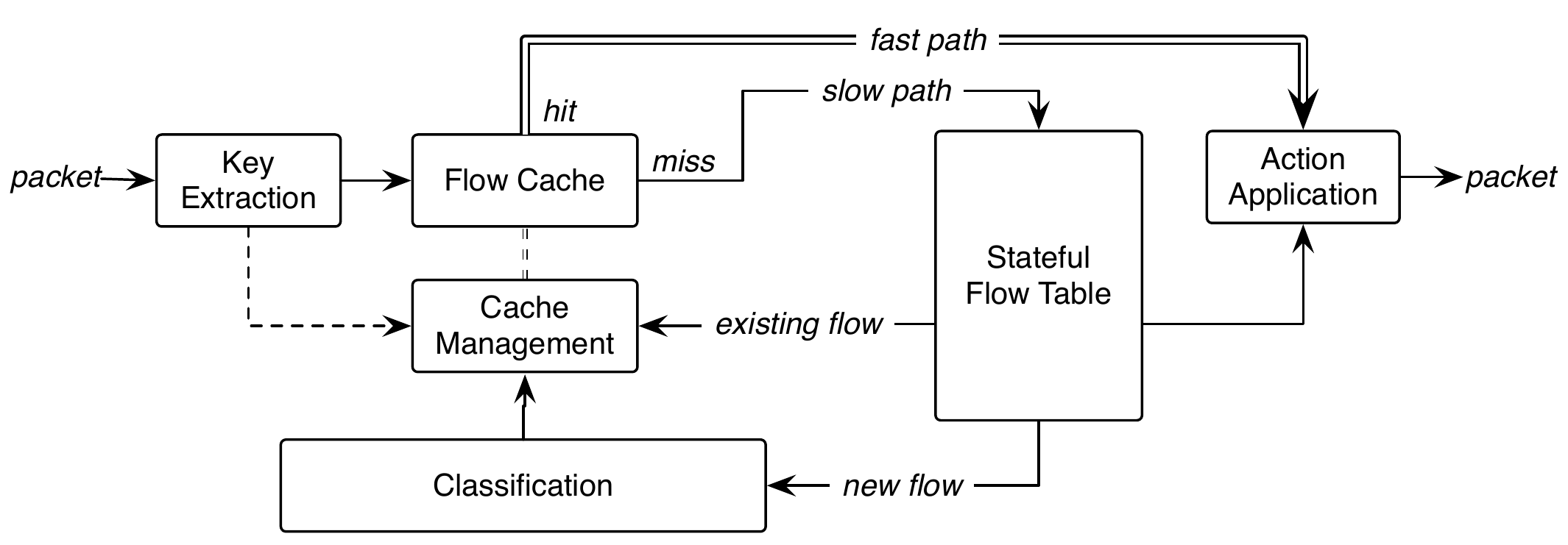} 
\caption{Flow Cache Management}
\label{fig:hp-flow-cache}
\end{figure}

Figure \ref{fig:hp-flow-cache} shows a classification pipeline focused around a flow cache and a backing stateful flow table.
It is useful to reiterate that a stateful flow table often resides in off-chip memory due to the number of concurrent flows tracked.
Stateful flow tables are managed by the control plane and a cache is often instantiated in the data plane to meet performance requirements.
There are two high-level scenarios encountered in stateful flow table management:
\begin{itemize}
\item \textbf{New Flow:} Requires classification and a new reservation in the stateful flow table.
\item \textbf{Existing Flow:} Track flow state in relevant table entry for associated packet.
\end{itemize}

A flow cache and its management algorithm must then contend with three scenarios triggered by incoming packets:
\begin{itemize}
\item \textbf{Compulsory Miss:} \emph{New flow}, not yet tracked in the stateful flow table.
\item \textbf{Capacity/Conflict Miss:} \emph{Existing flow}, not currently cached.
\item \textbf{Cache Hit:} \emph{Existing flow}, cached.
\end{itemize}

Cache management algorithms consist of two distinct components: the insertion and replacement policy.
A dynamic insertion policy may choose to bypass cache insertion if the flow is predicted to have a low probability of reuse before eviction.
In networking, bypassing flows based on static protocol heuristics is somewhat common, for example, ICMP messages.
However, there is an additional opportunity for an insertion algorithm to bypass tracked flows that have a low probability of cache reuse.
Both bypass and early replacement require a means to estimate reuse probability.

Additionally, a dynamic replacement policy may choose to evict a cache entry earlier than a static replacement policy otherwise would.
Since static replacement policies grant all entries equal opportunity for reuse, some flow entries are stuck waiting for eviction, taking up valuable cache resources even when the probability for reuse is markedly lower.
A dynamic replacement policy with an early eviction mechanism provides an additional opportunity to improve cache efficiency by removing stale entries sooner.

A packet without an associated flow table entry is considered a new flow, requiring classification.
Once the flow is classified, a flow table entry is reserved and the cache management algorithm may be consulted on whether to cache the entry (insert or bypass).
From the perspective of the flow cache, new flows triggering classification are compulsory misses -- an unavoidable slow-path.

Once a flow is tracked by the stateful flow table, the classification pipeline no longer needs to be consulted.
However, the effectiveness of the limited cache resources is ultimately up to the cache management algorithm.
This design aims to minimize the cache misses encountered by existing flows by exploring better cache management techniques.

Packets that encounter hits in the flow cache may simply update the cached flow table entry without needing to access off-chip flow table resources.
A more intelligent cache management algorithm has the potential to significantly improve the flow cache hit rate and thus data plane processing performance. 




The remaining section outlines the design details of the flow correlator mechanism.
Apart from design-time feature selection, the hashed perceptron flow correlation approach does not require global weight training.
The hashed perceptron feedback mechanism resembles an online training technique, adjusting correlation tables dynamically at run-time.

The adaptation of hashed perceptron cache management as a flow correlation mechanism is described in Section \ref{sec:hp-hp-design}.
As with many machine learning approaches, selecting useful Features at design time is non-trivial and requires artful exploration.
Our approach to feature design and metrics are covered in Section \ref{sec:hp-selection}.
Since feature selection consumed a majority of our effort during design iteration, we developed an iterative optimization approach generically applicable to tuning Hashed Perceptron-based cache management.
Section \ref{sec:hp-information-gain} describes this methodical approach to approximating information gained by differential performance analysis.

\subsection{Classifier Metrics} \label{sec:hp-classifier-metrics}
Early in feature exploration, we heavily leveraged averages and standard deviations of both pure and mixed features to help gain insight into input sparsity.
However, simply analyzing input and output distributions alone isn't sufficient to distinguish predictive features.

Analysis of binary classification problems is well-studied, and applicable across many research domains.
The aptly named \emph{Confusion Matrix} is commonly used to summarize the behavior of binary-outcome systems.
Table \ref{tab:hp-confusion-matrix} depicts a common representation of a confusion matrix.


\begin{table}[H]
\centering
\begin{tabular}{lcccl}
 & \multicolumn{1}{l}{} & \multicolumn{2}{c}{\textit{Predicted}} &  \\ \cline{3-4}
 & \multicolumn{1}{c|}{} & \multicolumn{1}{c|}{Positive} & \multicolumn{1}{c|}{Negative} &  \\ \cline{2-4}
\multicolumn{1}{c|}{} & \multicolumn{1}{c|}{Positive} & \multicolumn{1}{c|}{\cellcolor{pastelG}True Positive} & \multicolumn{1}{c|}{\cellcolor{pastelR}False Negative} &  \\ \cline{2-4}
\multicolumn{1}{c|}{\multirow{-2}{*}{\textit{Actual}}} & \multicolumn{1}{c|}{Negative} & \multicolumn{1}{c|}{\cellcolor{pastelR}False Positive} & \multicolumn{1}{c|}{\cellcolor{pastelG}True Negative} &  \\ \cline{2-4}
\end{tabular}
\caption{Confusion Matrix}
\label{tab:hp-confusion-matrix}
\end{table}

While simple ratio metrics are heavily used in certain domains, they aren't sufficiently robust for this work.
\emph{Accuracy} (Eq. \ref{eq-accuracy}) considers both positive and negative cases; however it is easily swayed by even moderately unbalanced datasets.
The \emph{F1} score (Eq. \ref{eq-f1}) is similarly unsuited as it is essentially \emph{accuracy} focused only on the positive set -- used primarily as a positive detection metric.


\begin{equation} \label{eq-accuracy}
accuracy = \frac{TP+TN}{TP+TN+FP+FN}
\end{equation}

\begin{equation} \label{eq-f1}
F1 = \frac{2 TP}{2 TP+FP+FN}
\end{equation}

Matthew's Correlation Coefficient (MCC) is a particularly useful metric to compare binary classifier performance \cite{matthews1975}.
MCC (Eq. \ref{eq-mcc}) is derived from Pearson's correlation coefficient applied specifically to binary classification.
MCC provides a balanced view of the confusion matrix, reducing decision bias from skewing the perceived accuracy.


\begin{equation} \label{eq-mcc}
MCC = \frac{TP*TN - FP*FN}{\sqrt{(TP+FP)(TP+FN)(TN+FP)(TN+FN)}}
\end{equation}


As a correlation coefficient, MCC is bound between $-1$ and $1$. 
A positive MCC indicates a correlation to correct predictions; while a negative MCC indicates anti-correlation.
An MCC near zero indicates a weak correlation, hinting at a poor prediction confidence resembling a random process.
As MCC approaches $\pm 1$, the predictor is deemed to have relatively higher prediction confidence.
Ultimately MCC provides a balanced view of binary classifier performance, particularly useful when operating within reasonable bias limitations \cite{chicco-mcc, Zhu-mcc-imbalance}.

\begin{table*}[b]
\centering
\begin{tabular}{lcccl}
 & \multicolumn{1}{l}{} & \multicolumn{2}{c}{\textit{Predicted}} &  \\ \cline{3-4}
 & \multicolumn{1}{c|}{} & \multicolumn{1}{c|}{Active} & \multicolumn{1}{c|}{Dormant} &  \\ \cline{2-4}
\multicolumn{1}{c|}{} & \multicolumn{1}{c|}{Active} & \multicolumn{1}{c|}{\cellcolor{pastelG} Active Correct (TP) } & \multicolumn{1}{c|}{\cellcolor{pastelR} Dormant Incorrect (FN) } &  \\ \cline{2-4}
\multicolumn{1}{c|}{\multirow{-2}{*}{\textit{Actual}}} & \multicolumn{1}{c|}{Dormant} & \multicolumn{1}{c|}{\cellcolor{pastelR} Active Incorrect (FP) } & \multicolumn{1}{c|}{\cellcolor{pastelG} Dormant Correct (TN) } &  \\ \cline{2-4}
\end{tabular}
\caption{Flow Correlation Confusion Matrix}
\label{tab:hp-confusion-matrix-2}
\end{table*}

To assist in subsequent sections, Table \ref{tab:hp-confusion-matrix-2} specializes in the confusion matrix around predictions of flow activity in the context of cache management.
Section \ref{sec:hp-reinforcement} will connect each case in the confusion matrix to the hashed perceptron training algorithm.

\subsection{Flow Correlator Design} \label{sec:hp-hp-design}


The hashed perceptron technique provides a structured mechanism to leverage multiple prediction tables for improved overall accuracy and resiliency.
The hashed perceptron training algorithm plays a crucial role in keeping these predictor tables balanced across multiple features.
Apart from design-time feature selection, the hashed perceptron flow correlation approach does not require global weight training.
The hashed perceptron feedback mechanism resembles an online training technique, adjusting correlation tables dynamically at run-time.

The hashed perceptron consists of correlation tables (feature tables) consulted during inference.
Each correlation table is indexed by a feature and contains saturating counters (weights) resembling typical prediction tables.
The Hashed Perceptron differs from single-perspective predictors by combining the predictions from multiple feature tables to produce a final combined decision.
The magic of the hashed perceptron centers around the ability to manage multiple tables cohesively using a prediction feedback mechanism (the hashed perceptron training algorithm).

Flow table cache management relies on accurately predicting reuse for a given flow.
This ultimately boils down to predicting packet inter-arrival patterns for each active Flow.
More specifically, predicting the likelihood of a subsequent packet arrival for a given flow within the relative time window allotted by the cache's working set.


The Hashed Perceptron mechanism is conceptually split into two phases: inference and reinforcement.
Section \ref{sec:hp-inference} describes the process of generating a reuse prediction for every packet arrival.
Section \ref{sec:hp-reinforcement} describes the Hashed Perceptron training algorithm as the basis for maintaining prediction tables.

\subsubsection{Inference} \label{sec:hp-inference}
Cache management predictions are generated in real-time by the flow correlator inference mechanism.
Every packet event triggers an inference prediction for cache management, thus it is crucial that inference is able to meet the packet processing requirement of the underlying design target.

Cache management handles two scenarios depending on the cached state of the flow entry.
First, a flow cache miss triggers a cache management decision to insert or bypass the flow entry.
Secondly, cache management seeks to identify optimal scenarios for early eviction.
An early eviction, or conversely a reuse prediction, is performed on a flow cache hit.
In both scenarios, cache management consults flow correlation inference to estimate the likelihood of flow entry reuse relative to overall cache pressure.

Both predictions are essentially re-framing an underlying flow entry reuse prediction around a relative ranking of flow activity.
Thus, the flow correlator inference and training pipelines are organized around the concept of \emph{active} or \emph{dormant} flows.
Both bypass and early eviction share table resources, interpreting positive table weights as likely to be \emph{active} and negative weights as likely \emph{dormant}.

\begin{figure}[htb]
\centering
\includegraphics[width=0.8\linewidth]{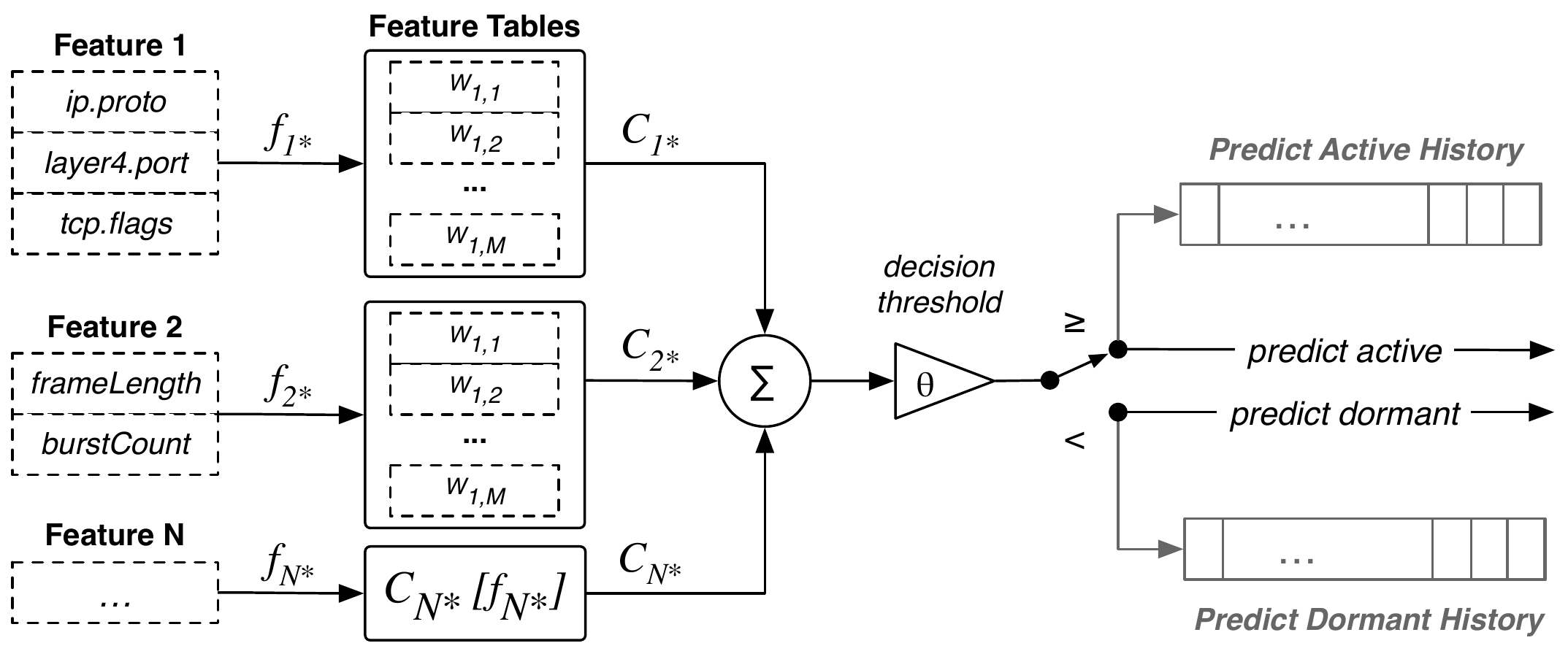} 
\caption{Flow Correlator Inference}
\label{fig:hp-inference}
\end{figure}

Figure \ref{fig:hp-inference} outlines the flow correlator's inference pipeline.
In order for cache management to generate an inference, the feature components must first be gathered from the packet header bitfields and the relevant flow table entry.
Stateless protocol feature components originate from the packet headers and are readily available by preexisting classification \emph{Key Extraction} operations.
Stateful feature components (e.g. burstCount) are pulled from a cached flow table entry during a cache hit, or simply zero during a cache miss.

Each assembled feature ($f_{N*}$) indexes into a corresponding feature table providing a mapping to independent correlations ($C_{N*}$).
The decision threshold ($\theta$) is applied to the sum of each feature's correlation.
Since $\theta$ is zero for this flow correlator design, the magnitude of the accumulated correlations can be interpreted as confidence and the sign represents the decision.


On inference, each feature contributes an opinion as a correlation count.
The hashed perceptron structure accumulates the correlations from all features, thresholding to create a reuse prediction.
The confidence of the feature's opinion can be inferred by the magnitude of the correlation count.
Similarly, the confidence of the overall prediction is the magnitude of all accumulated correlations.



Feature tables map a disjoint input feature-space to cache reuse correlations.
Loosely resembling a hash, features are often orthogonal combinations of several bit fields.
These feature table lookups are performed in parallel.
Each feature table holds saturating correlation counters, commonly referred to as weights.
Each feature table's length is $2^{m}$ entries where $m$ refers to the bit-width of the corresponding feature assembly.
The implementation requirements of each feature are then simply $w * 2^{m}$ bits, where $w$ refers to the saturating counter bit-width.


Flow entries marked for early eviction by the flow correlation mechanism are in the cache, but prioritized for replacement.
When a flow entry needs to be inserted and no entries are marked for early eviction, replacement falls back to LRU.
If a flow entry marked for early replacement encounters a hit before eviction, the prediction is corrected and a training event is triggered.

\subsubsection{Reinforcement} \label{sec:hp-reinforcement}
The hashed perceptron mechanism consists of an online training algorithm organized around prediction feedback.
Applied to flow correlation, prediction feedback seeks correlations of flow entry reuse across feature vectors.
With respect to the available cache resources, the flow correlation predictor aims to rank flow entries by the likelihood of reuse within the available cache working set.



\emph{Active} flow predictions are inserted into the active history queue, while \emph{dormant} flow predictions are inserted in the dormant history queue.
In order to reinforce cache management decisions, these feedback queues must be searchable by a unique flow identifier.
The flow identifier implementation leveraged by the flow cache may be borrowed for this purpose.

\begin{figure}[htb]
\centering
\includegraphics[width=0.6\linewidth]{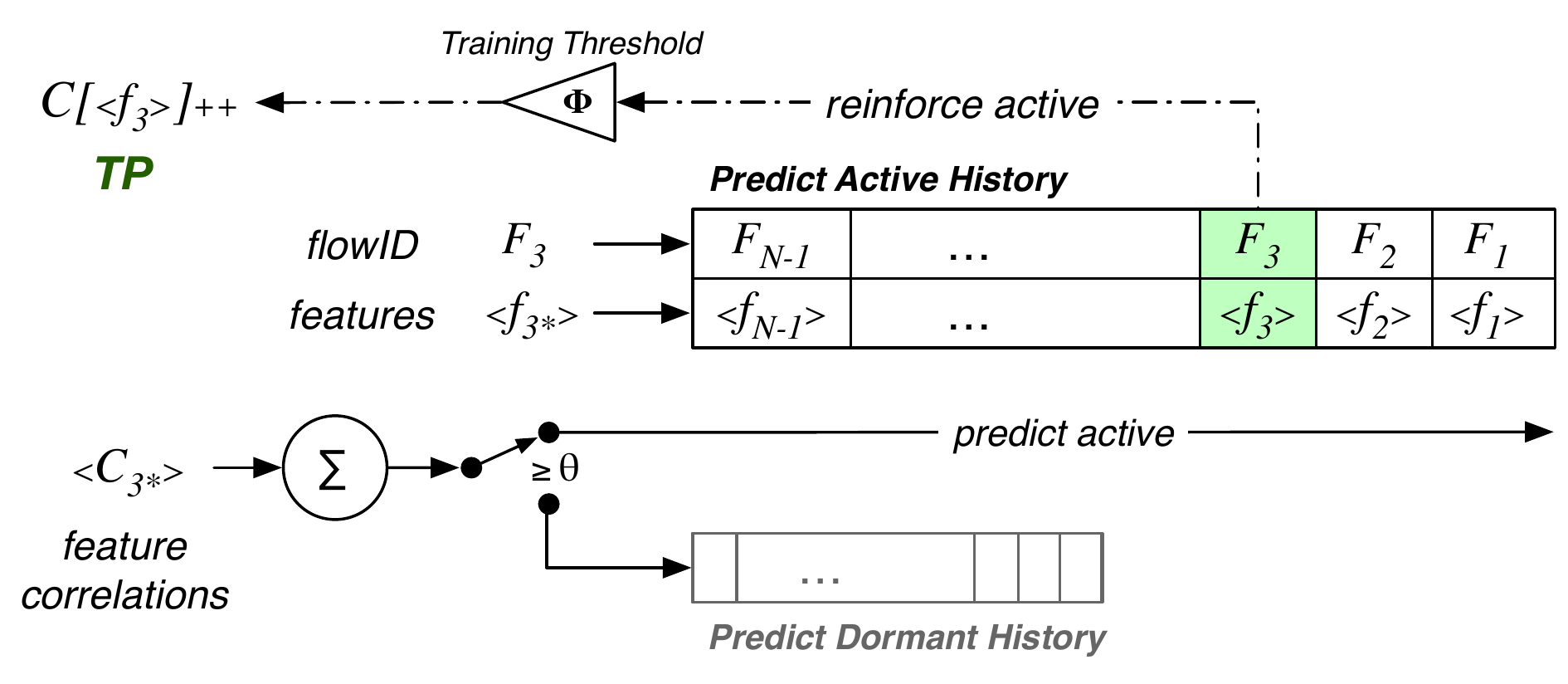} 
\caption{Active Correct (TP) Reinforcement}
\label{fig:hp-training-tp}
\end{figure}
\begin{figure}[htb]
\centering
\includegraphics[width=0.6\linewidth]{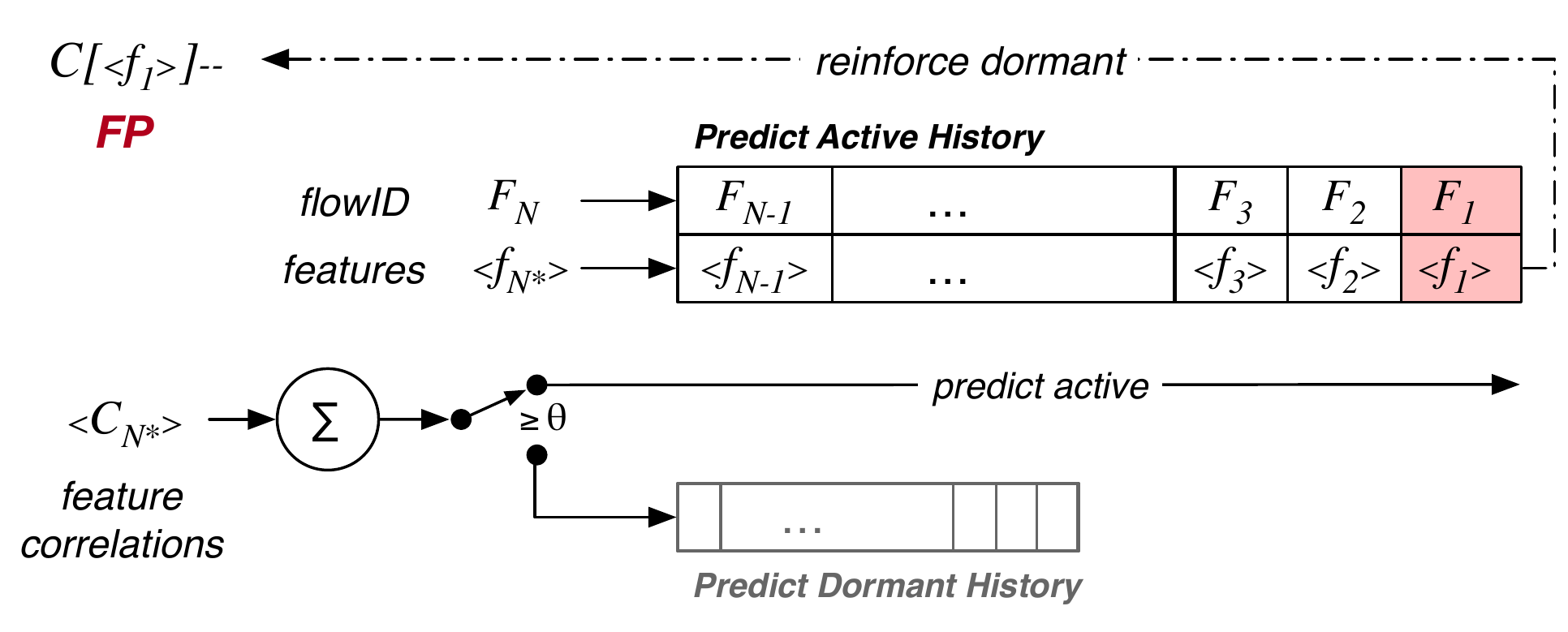} 
\caption{Active Incorrect (FP) Training}
\label{fig:hp-training-fp}
\end{figure}

Hashed perceptron training reinforcement consists of four potential scenarios, each corresponding to entries in the confusion matrix.
\emph{Active} prediction reinforcement is shown in Figure \ref{fig:hp-training-tp} and Figure \ref{fig:hp-training-fp} covering the \emph{true positive} and \emph{false positive} cases, respectively.

In Figure \ref{fig:hp-training-tp}, the example flow $F_{3}$ has a prior \emph{active} flow prediction awaiting confirmation in the \emph{active history} queue.
Since the prior reuse prediction is now proved correct, the feature vector, $\langle f_{3} \rangle$, associated with the previous prediction is popped off of the queue.
If the prediction confidence associated with $\langle f_{3} \rangle$ do not surpass the training threshold, the corresponding feature table correlation counts are incremented.
The new reuse prediction associated with $F_{3}$ is also pushed into the \emph{active history} queue along with the associated new feature vector, $\langle f_{3*} \rangle$.

Figure \ref{fig:hp-training-fp} outlines how incorrect \emph{active} flow predictions are determined.
Example flow $F_{N}$ does not have a previous reuse prediction, yet is predicted active by inference.
In order to free a slot, the oldest prediction is popped off the end of the \emph{active history} queue and considered incorrect.
The example flow $F_{1}$ and feature table correlation counts associated with feature vector $\langle f_{1} \rangle$ are decremented.
Note the training threshold is omitted since the \emph{active} flow prediction associated with $F_{1}$ was determined to be incorrect.
The new reuse prediction associated with $F_{N}$ is pushed into the \emph{active history} queue along with its feature vector, $\langle f_{N*} \rangle$.

The next two figures illustrate \emph{dormant} flow prediction reinforcement covering the \emph{true negative} and \emph{false negative} cases of the confusion matrix, respectively.
Conversely to \emph{active history}, entries popped off the end of the \emph{dormant history} queue are considered correct.

\begin{figure}[htb]
\centering
\includegraphics[width=0.6\linewidth]{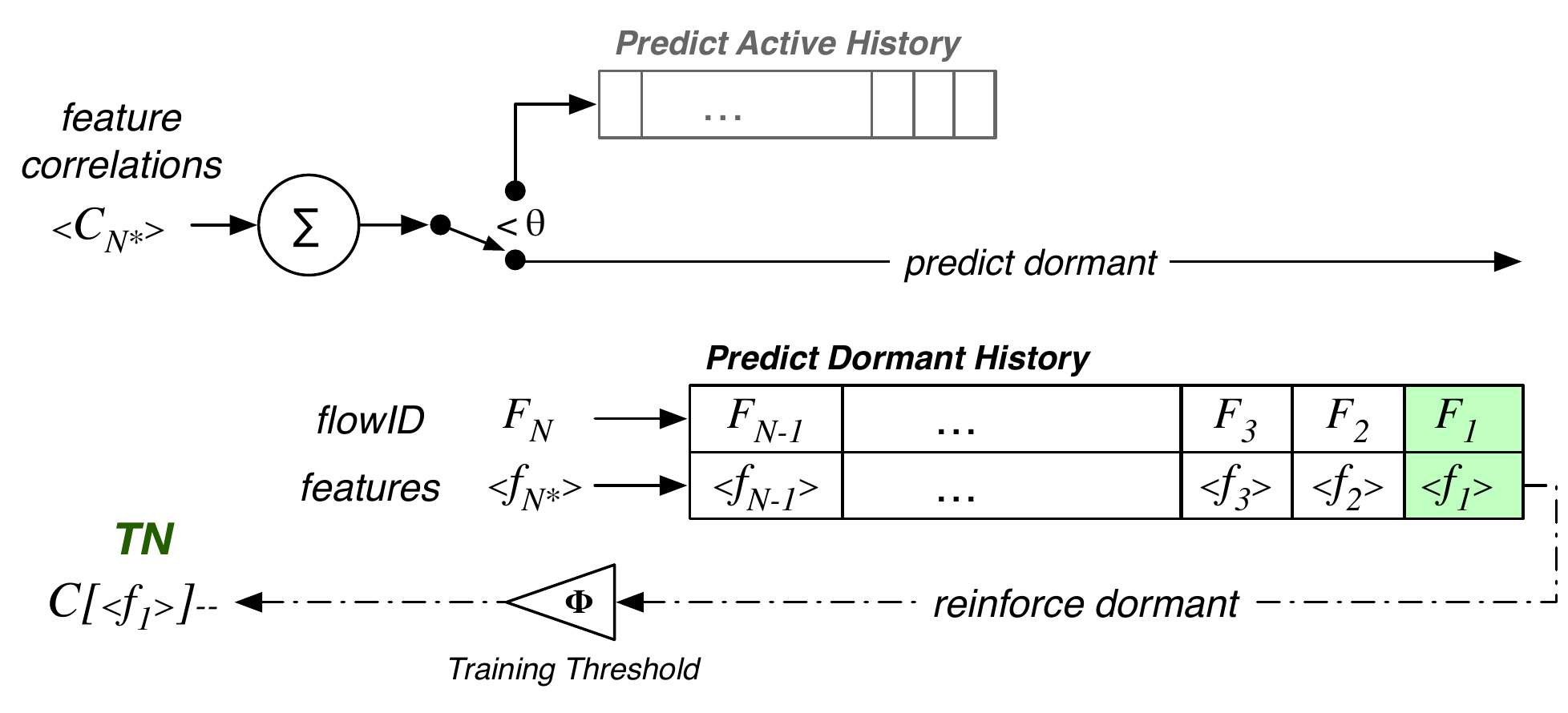} 
\caption{Dormant Correct (TN) Reinforcement}
\label{fig:hp-training-tn}
\end{figure}

Figure \ref{fig:hp-training-tn} illustrates example flow $F_{N}$ predicted \emph{dormant} during inference.
Finding no prior prediction associated with $F_{N}$, the oldest prediction in the \emph{dormant history} queue is popped off the end.
If the prediction confidence associated with $\langle f_{1} \rangle$ does not exceed the training threshold, the corresponding feature table correlation counts are decremented -- reinforcing the correct \emph{dormant} prediction.
The new \emph{dormant} prediction associated with $F_{N}$ is pushed into the \emph{dormant history} queue along with its feature vector, $\langle f_{N*} \rangle$.

\begin{figure}[htb]
\centering
\includegraphics[width=0.6\linewidth]{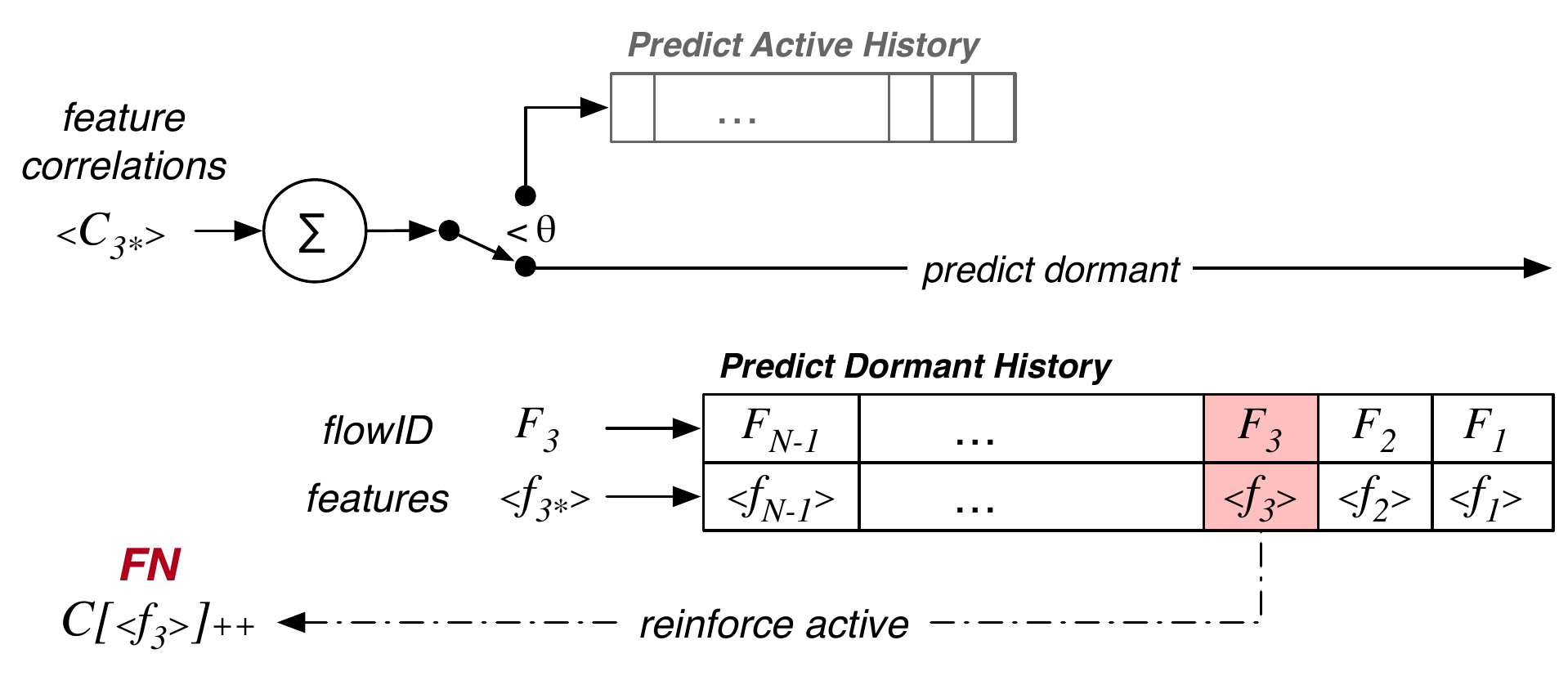} 
\caption{Dormant Incorrect (FN) Training}
\label{fig:hp-training-fn}
\end{figure}

In Figure \ref{fig:hp-training-fn}, the example flow $F_{3}$ has a prior \emph{dormant} flow prediction awaiting confirmation in the \emph{dormant history} queue.
The prior \emph{dormant} prediction was caught before falling off the end of the queue, indicating that feature vector $\langle f_{3} \rangle$ may contain hints towards flow reuse.
The existing $F_{3}$ entry and associated feature vector, $\langle f_{3} \rangle$, is removed from the queue and the respective feature table correlation counts are incremented.
Note once again that the training threshold is omitted since the \emph{dormant} flow prediction associated with $F_{3}$ was determined to be incorrect.
Finally, the new reuse prediction associated with $F_{3}$ is pushed into the \emph{dormant history} queue along with the associated new feature vector, $\langle f_{3*} \rangle$.


Seznec developed a dynamic training thresholding mechanism for the Geometric History Length branch predictor \cite{seznec-thresholding}.
We leveraged Seznec's adaptive mechanism for the \emph{training threshold} ($\Phi$), which also requires the \emph{decision threshold} ($\theta$) to be centered around zero.
Through experimentation, we confirmed that Seznec's target $1\!:\!1$ ratio of correct to incorrect predictions resulted in the best overall predictor accuracy, matching our experimentally tuned parameters.

While the optimal $\Phi$ depends on the cache pressure, we observed a notably stable threshold once the predictor warmed up and aggressiveness balanced.
It is also notably advantageous to keep the decision threshold ($\theta$) intuitively at zero -- preserving the numeric symmetry of weights as well as accumulated correlations.


\subsection{Feature Design} \label{sec:hp-selection}
Table \ref{tbl:hp-all-features} lists features explored during the design process with remarks on the approximate intent behind each combination.
There is limited information available from fields contained within each packet.
However, we found notable value in associating temporal patterns with hints from packet fields.
The features outlined in this table are nowhere near an exhaustive exploration, but rather a first pass to identify useful combinations.

\begin{table*}[t]
\centering
\singlespacing
\begin{tabular}{clll}
\textbf{$f_{\#}$} & \textbf{Feature Type} & \textbf{Feature Components} & \textbf{Remarks} \\
0  & Control & $UniformRandom$ & Uncorrelated: map to random entry \\
1  & Pure Mix & $ipProto \wedge min(srcPort, dstPort)$ & Protocol hints \\
2  & Pure Mix & $ipv4Dst[31\!\colon\!16] \wedge dstPort$ & Destination service \\
3  & Pure Mix & $ipv4Src[31\!\colon\!16] \wedge srcPort$ & Source service \\
4  & Pure & $flags$ & TCP/IP flags \\
5  & Pure & $srcPort \wedge dstPort$ & Bi-directional TCP/UDP ports \\
6  & Pure Mix & $f_{TCP}\!\ll\!7 \wedge f_{1}$ & Mix TCP flags with protocol hint \\
7  & Pure & $(ipv4Dst \wedge ipv4Src)[31\!\colon\!16]$ & Bi-directional upper 16-bits IP addresses \\
8  & Pure & $(ipv4Dst \wedge ipv4Src)[23\!\colon\!8]$ & Bi-directional middle 16-bits IP addresses \\
9  & Pure & $(ipv4Dst \wedge ipv4Src)[15\!\colon\!0]$ & Bi-directional lower 16-bits of IP addresses \\
10 & Pattern & $\lceil flowPackets \rceil_{16}$ & Packets since start of flow \\
11 & Pure & $ipLength$ & IP frame length \\
12 & Pure & $flowID$ & 5-tuple flow identifier \\
13 & Pure Mix & $f_{12} \wedge f_{11}$ & Unroll frame length over flowID \\
14 & Pattern & $\lceil refCount \rceil_{16}$ & Hit count while inserted into cache \\
15 & Pattern & $\lceil burstCount \rceil_{16}$ & Hit count while in MRU position \\
16 & Pattern Mix & $\{ \lceil f_{14} \rceil_8, \lceil f_{15} \rceil_8 \}$ & Concatenate RefCount and BurstCount \\
17 & Pure + Pattern & $f_{16} \wedge f_{12}$ & Unroll Burst/RefCount over flowID \\
18 & Pure + Pattern & $f_{16} \wedge f_{7}$ & Unroll Burst/RefCount over upper IP \\
19 & Pure + Pattern & $f_{16} \wedge f_{8}$ & Unroll Burst/RefCount over middle IP  \\
20 & Pure + Pattern & $f_{16} \wedge f_{9}$ & Unroll Burst/RefCount over lower IP \\
21 & Pure + Pattern & $f_{11} \wedge f_{7}$ & Unroll frame length over upper IP \\
22 & Pure + Pattern & $f_{11} \wedge f_{8}$ & Unroll frame length over middle IP  \\
23 & Pure + Pattern & $f_{11} \wedge f_{9}$ & Unroll frame length over lower IP \\
24 & Pure Mix & $ipFragOffset \wedge f_{4} \wedge f_{11} \wedge f_{12}$ & \makecell[l]{Unroll fragment offset, flags, and \\frame length over flowID} \\
25 & Pure Mix & $f_{7} \wedge f_{4}$ & Unroll flags over upper IP \\
26 & Control & $NULL$ & Sparsity: map to single entry\\
27 & Pure + Pattern & $\{f_{11}, \lceil f_{15} \rceil_8 \}$ & Concatenate BurstCount with frame length \\
28 & Pure + Pattern & $f_{16} \wedge f_{11}$ & Unroll Burst/RefCount over frame length
\end{tabular}
\longcaption{Complete List of Explored Features}{
Concatenation syntax is represented as $\{A, B\}$.
$\lceil A \rceil_N$ refers to the ceiling function: $min(A, N)$.
The XOR operation is indicated by $\wedge$.
$A\!\ll\!N$ represents left shift left by N-bits.
Bit-field selection is indicated using the syntax $[M\!\colon\!N]$, where bit positions $M$ through $N$ are extracted.}
\label{tbl:hp-all-features}
\end{table*}


The \emph{Feature Type} column approximately categorizes each feature by information sources.
There are two potential sources of information: \emph{Pure} (stateless) and \emph{Pattern} (stateful).
Pure feature components originate from packet headers.
Pattern feature components introduce temporal information, not already available within packet headers.
\emph{Pure Mix} are simple combinations of packet header bit-fields.
\emph{Pattern Mix} refers to combinations of temporal states.
Finally, \emph{Pure + Pattern} are combinations of both packet header bit-fields and temporal state.


\subsubsection{Pure Features} \label{sec:hp-pure-features}
There are really only five pure feature components originating from packet headers explored in this study.
These include IP source and destination addresses, IP protocol identifier, TCP/UDP port numbers, IP frame length, and few grouped IP/TCP flags.
The IP flags assembled in $f_{4}$ are: DF and MF.
The TCP flags assembled in $f_{4}$ and $f_{6}$ are: SYN, FIN, RST, PSH, ACK, URG. 


\subsubsection{Temporal Pattern Features} \label{sec:hp-pattern-features}
Pattern features are distinguished from Pure features as they require temporal metadata in the flow table entry.
These pattern feature components do not originate from packet headers but offer the ability to associate sequences or temporal patterns between packets.
Pattern-based features proved to be valuable sources of information but required mixing to unlock their potential.

The flowPackets feature ($f_{10}$) is simply the packet count since the flow started, a counter often already present in stateful flow tables.
RefCount ($f_{14}$) and BurstCount ($f_{15}$) on the other hand are similar counters, but unique to the lifetime of the cached flow table entry.
RefCount is simply the packets observed since the flow was inserted into the cache, while BurstCount is the packets observed while the entry remains in the MRU position.

\subsubsection{Combined Features}
There are endless ways to assemble features with complementary pure and pattern-based components.
There were two conceptual approaches when attempting to constructively combine feature components: \emph{orthogonal combination} and \emph{correlation unrolling}.

\emph{Orthogonal combinations} allows for better utilization of feature table resources through a union of independent feature components.
For example some feature components are useful for bypass predictions, while others provide hints toward early eviction.
Usually, we observed a trade-off in feature information density at the cost of collisions and reduced confidence.
However, we also observed the potential for coincidental benefit -- correlations simultaneously occurring, strengthening feature confidence.

\emph{Feature unrolling}, as an approach towards enabling constructive feature combinations aiming to distribute global correlations across subsets of flow designators such that opportunities for independent correlations may arise.
For example, temporal patterns tend to perform better when associated with partial IP address subsets, protocol, or port designators.
Correlation unrolling is a balance and can either benefit or detract from feature performance.

Pattern-based features are particularly complex in the context of networking.
Subsets of flows may exhibit similar patterns, but all flows together tend to be quite noisy with conflicting inter-arrival patterns.
Thus, we observed notable utility in correlating pattern-based feature components across protocol-differentiating components.
It is also important to conceptualize that each packet in a given flow maps to different flow table entries throughout the sequence.
The inter-arrival access patterns is thus encoded across multiple entries in any given table.

There is certainly a trade-off ranging from targeting independent per-flow correlations to strategic groupings sharing correlator entries.
Unlike bloom filter designs, where hashes are desired to be as uniquely identifying and free of collisions as possible, 
there is a balance in hashed perceptron feature construction where carefully selected subsets to improve correlation potential.

The designer can choose to combine orthogonal features or unroll features by spreading potential correlations out over subsets of flows.
Since necessary unrolling can increase a feature's effective training latency, there is a non-obvious trade-off between the concept of sharing correlations across subsets of flows as well as enabling flows to maintain independent correlations.
The features presented in this work are almost certainly non-optimal but do provide insight towards successful (as well as unsuccessful) combinations.


\subsubsection{Control Features} \label{sec:hp-control-features}
To help understand the natural bias of the system, two control features were included in the study: UniformRandom ($f_{0}$) and NULL ($f_{26}$).
These were only used in the feature exploration process as a likeness comparison
UniformRandom simply chooses a table entry randomly based on a uniform random distribution, providing insight into the hysteresis of the system.
NULL is effectively a single table entry (5-bit saturating counter), representing sparse feature behavior.

Combining too many components reduces the opportunities for correlations.
We have observed that heavily mixed features, such as features that combine the full flow identifier (flowID, $f_{12}$) tend to under-perform features that combine a partial flow identifier such as $f_{7}$.
All components mixed together into a single feature tends to approach UniformRandom ($f_{1}$) while sparse features tend to resemble NULL ($f_{26}$).

\subsection{Feature Metrics} \label{sec:hp-feature-metrics}
Since the hashed perceptron structure consists of multiple independent prediction tables contributing opinions towards an overall prediction, there is a need to analyze feature performance relative to the overall system.
Just as MCC is a notably useful classifier metric, it can also be calculated with respect to each feature, independently.

The Hashed Perceptron training feedback is assumed to be ground truth.
Absolute truth is non-trivial to ascertain as each misprediction has implications that propagate forward in time.
Imperfect, ground truth is a practical compromise that extends to all feature analysis techniques in this work.




\subsubsection{Feature Influence} \label{sec:hp-influence}

While MCC is well suited for relative feature comparisons, it doesn't quite grant insight into the confidence of predictions.
Specifically in the context of the hashed perceptron mechanism, each feature has the potential to influence the system prediction equally.
However well behaved features self-moderate their confidence by the weight of their contribution.
Understanding how a feature contributes to the overall system provides valuable feedback into the behavior of feature components as well as intuition into successful combinations.

\begin{table}[H]
\centering
\begin{tabular}{lcccl}


 & \multicolumn{1}{l}{} & \multicolumn{2}{c}{\textit{Predicted}} &  \\ \cline{3-4}
 & \multicolumn{1}{c|}{} & \multicolumn{1}{c|}{Positive} & \multicolumn{1}{c|}{Negative} &  \\ \cline{2-4}
\multicolumn{1}{c|}{} & \multicolumn{1}{c|}{Positive} & \multicolumn{1}{c|}{\cellcolor{pastelG} $\frac{\sum \vec{w}_f }{TP}$ } & \multicolumn{1}{c|}{\cellcolor{pastelR} $\frac{\sum \vec{w}_f }{FN}$ } & \\[1mm] \cline{2-4}
\multicolumn{1}{c|}{\multirow{-2}{*}{\textit{Actual}}} & \multicolumn{1}{c|}{Negative} & \multicolumn{1}{c|}{\cellcolor{pastelR} $\frac{- \sum \vec{w}_f}{FP}$ } & \multicolumn{1}{c|}{\cellcolor{pastelG} $\frac{- \sum \vec{w}_f}{TN}$ } &  \\[1mm] \cline{2-4}
\\


\end{tabular}
\caption{Influence Matrix}
\label{tab:hp-influence-matrix}
\end{table}

Table \ref{tab:hp-influence-matrix} shows how influence is an extension of the confusion matrix.
The influence matrix is categorized with respect to the overall correctness of the system prediction (TP, TN, FP, FN).
The vector notation simply indicates that feature weights ($\vec{w_f}$) are accumulated independently, with respect to each system prediction.

Note that influence of a single-feature predictor is simply the average weight with respect to that predictor's confusion matrix.
However, the influence metric becomes more than just an average weight when per-feature contributions are tracked with respect overall system decisions.
Influence grants insight into the feature's contribution, or lack thereof, relative to the correctness of the system's overall prediction.

Influence can be analyzed with respect to each quadrant of the confusion matrix.
As noted above, the influence vector notation $\vv{I\!N\!F}$ indicates individual per-feature weight accumulations ($\vec{w}_f$) with respect to system decisions.
To simplify analysis, feature influence can be grouped with respect to correct and incorrect system predictions.

\begin{equation}
\vv{I\!N\!F}_{correct} = 
\frac{\displaystyle\sum^{T\!P} \vec{w}_f}{TP} - 
\frac{\displaystyle\sum^{T\!N} \vec{w}_f}{TN}
\label{eq:f-correct-inf}
\end{equation}

Equation \ref{eq:f-correct-inf} defines \emph{Correct Influence} as the contributed weight towards correct decisions.
The numerical value of correct influence is the average weight contributed towards correct predictions.
Correct Influence can be interpreted as a feature's confidence when deemed correct, providing unique insight alongside prediction accuracy.

\begin{equation}
\vv{I\!N\!F}_{incorrect} = 
\frac{\displaystyle\sum^{F\!N} \vec{w}_f}{FN} - 
\frac{\displaystyle\sum^{F\!P} \vec{w}_f}{FP}
\label{eq:f-incorrect-inf}
\end{equation}

Similarly, Equation \ref{eq:f-incorrect-inf} defines the \emph{Incorrect Influence} as the contributed weight towards incorrect decisions.
Incorrect influence can be interpreted as a feature's overconfident contribution towards incorrect predictions.
Overconfident features ultimately detract from an otherwise potentially correct prediction, a notably useful insight unique to influence.
To improve interpretation, \emph{Incorrect Influence} maintains the same directionality alongside \emph{Correct Influence}, where weight contributions towards incorrect predictions are negative numeric values, while correct contributions despite being overruled are positive.


The negation of the \emph{FP} and \emph{TN} accumulations map all correct predictions into a positive influence and incorrect predictions into negative influence.
Total influence can them be considered the sum of correct and incorrect influence as shown in Equation \ref{eq:f-total-inf}.

\begin{equation}
\vv{I\!N\!F}_{total} = \vv{I\!N\!F}_{correct} + \vv{I\!N\!F}_{incorrect}
\label{eq:f-total-inf}
\end{equation}


Ideally, features would remain impartial until sufficient correlation occurs to form a reliable prediction.
In practice, features can be confidently wrong or even marginally correct while still providing value to the system.
Independently, features often produce noisy and inaccurate predictions.
Considered together with the hashed perceptron structure, feature noise is averaged away while simultaneous correlations combine to improve prediction quality.

The notable difference between MCC and Influence centers around the penalty attributed to mispredictions.
MCC treats all feature predictions equally, regardless of the accompanying confidence.
Rather, influence provides a perspective into the confidence towards correct decisions as well as overconfidence towards incorrect decisions.
MCC provides a robust balanced accuracy, where influence grants insight into contributions.

Features able to regulate their overconfidence are notably valuable to the system -- even if they only occasionally grant unique perspectives.
While there will inherently be inaccuracies in any history-based pattern correlation mechanism, understanding feature contributions towards correct/incorrect decisions helps during feature selection and tuning.

\subsubsection{Inferring Information Gain} \label{sec:hp-information-gain}
Information gain has been influential in decision trees \cite{decision-tree-ig} as a tool to estimate mutual information.
Decisions trees leverage information gain as a means to estimate mutual information between competing branches when growing as well as trimming decision trees.

Inspired by information gain, we leverage a similar technique comparing the differential performance gain (or loss) between adjacent simulation runs.
Algorithm \ref{alg:hp-sweep} defines a simulation sweep across an ordered set of features.
By incrementally adding features in ranked order, information gained by the added feature can be estimated through a resulting change in the simulation's performance metric.
Plotting the improvement (or degradation) of each additional feature grants insight into any objectively useful information added to the system.

\begin{algorithm}
  \caption{Simulation Feature Sweep}
  \label{alg:hp-sweep}
  \begin{algorithmic}[1]
    \Function{Sweep}{$\vec{R}$} \funclabel{alg:sweep}
    \Comment{Simulation hit-rate sweep across feature ranking}
      \For{$n \gets 1$ to $\Call{Length}{\vec{R}}$}
        \State $\vec{r}\gets \{\vec{R}[0],~...,~\vec{R}[n]\}$
            \Comment{Feature subset to include in simulation run}
        \State $\vec{P}[i]\gets \Call{Simulate}{\vec{r}}$
      \EndFor
      \State \Return $\vec{P}$
        \Comment{Simulation performance metric in inclusive ranked order}
      \EndFunction
  \end{algorithmic}
\end{algorithm}

This \emph{feature sweep} estimation is particularly useful as both MCC and influence are inherently unable to identify shared information between competing features.
Realizing that competing rankings can be compared by sweeping features,
 we stumbled on an approach to prune shared information.
 Algorithm \ref{alg:hp-ig-loop} describes our approach to formalize our process of feature ranking in the context of a hashed perceptron prediction system.

\begin{algorithm}
  \caption{Differential Information Gain}
  \label{alg:hp-ig-loop}
  \begin{algorithmic}[1]

    \Function{IG}{$\vec{F}$} \funclabel{alg:ig}
    \Comment{Inferring information gain through differential system improvement}
      \State $p_{0}\gets \Call{Simulate}{\vec{F}}$
      \State $\vec{R}_{0}\gets \Call{MCC}{p_{0}}$
        \Comment{Initial ranking provided by MCC}
      \State $i\gets 1$
      \While{$\vec{R}_{i} \not= \vec{R}_{i-1}$}
        \Comment{Repeat until ranking is stable}
        \State $\vec{p}\gets$ \Call{Sweep}{$\vec{R}_{i-1}$}
        \State $\vec{q}\gets \Call{AdjacentDifference}{\vec{p}}$
            \Comment{Calculate relative improvement gain}
        \State $\vec{R}_{i}\gets \Call{Sort}{\vec{q}}$
        \State $i\gets i + 1$
      \EndWhile\label{endloop}
      \State \Return $\vec{R}_{i}$
        \Comment{New ranking sorted by relative improvement}
      \EndFunction
  \end{algorithmic}
\end{algorithm}

Ranking features purely by information gain is tempting, but unfairly favors the features already selected over the differential new feature added.
This algorithm greedily optimizes to the nearest local minimum.
However, this technique is notably useful in eliminating mutual information from an already decent initial feature ranking provided by MCC.

\section{Analysis} \label{sec:hp-analysis}
The primary evaluation goal of this work is to assess the viability of applying the Hashed Perceptron correlation mechanism to stateful flow table cache management.
Contrasted with the cache optimality limit study in Section \ref{sec:hp-opt}, this feasibility study provides insight into  practically capturable locality.



\subsection{Methodology} \label{sec:hp-methodology}
The simulator developed for this study allowed for a quick exploration of cache management strategies in the context of stateful network flow tables.
This work focused on cache hit rate as the primary performance metric with the expectation that cache hit rate translates directly into improved throughput and ultimately cache efficiency.
While this cache hit-rate simulation does not model the latency intricacies incurred by cache misses as well as flow table management, it does showcase the relative prediction accuracy of competing cache management algorithms.
Focusing on cache hit-rate simulation for the viability study allows for quick design iteration.

\subsubsection{Network Traffic Datasets}
It is tempting to leverage synthesized network traffic based on statistical models of packet arrival behavior \cite{traffic-gen, traffic-gen-validation}.
However, the random packet inter-arrival assumptions of these models would both hinder the cache management algorithm as well as introduce uncertainty around the accuracy of the synthesized traces.
While these models can quite accurately resemble real inter-arrival flow behavior, actual packet arrival behavior is inherently complex and beyond our confidence to model.
This viability study hinges on having representative datasets of network traffic to validate the premise of capturable temporal locality through more advanced cache management techniques.

For this study, we chose to leverage network packet capture (PCAP) traces rather than synthetic traffic in order to preserve temporal inter-arrival patterns theorized to be present in Section \ref{sec:hp-patterns}.
CAIDA's offers passive PCAP traces within their Equinix internet exchange\footnote{Internet Exchange Points (IXPs) are crucial infrastructural components of the Internet where network operators (peers) exchange traffic.}
data centers. 
We leveraged PCAP traces from CAIDA Equinix Sanjose, Chicago, and NYC datasets spanning 2012 to 2018 \cite{caida-cache}.
Traffic flowing through these exchange points should contain a representative sampling network traffic commonly found on the Internet.
CAIDA's Equinix datasets were chosen primarily because of the availability and diversity of the traffic provided and available to researchers.

\begin{table}[ht]
\small
\centering
\singlespacing
\begin{tabular}{|r|l|}
\hline
\textbf{CAIDA Dataset} & 
    Equinix Sanjose January 2012 \\ \hline
\textbf{Interface} &
    10 Gbps link (6 Gbps average, bi-directional) \\ \hline
\textbf{Packet Rate} & 
    1M / second \\ \hline
\textbf{New Flows} & 
    30k / second \\ \hline
\textbf{Active Flows} & 
    300k over 4-second window \\ \hline
\textbf{Trace Length} & 5 minutes \\ \hline
\end{tabular}
\caption{Sample of CAIDA Packet Arrival Statistics}
\label{tbl:hp-caida-stats}
\end{table}

The CAIDA Equinix PCAP traces utilized for this study are captured on 10 Gbps links and taken over one-hour intervals, split into one-minute files.
Table \ref{tbl:hp-caida-stats} provides a summary of packet arrival and flow turnover statistics.
The CAIDA PCAP files contain pseudo-anonymized network layer IPv4 addresses, preserving subnets while anonymizing exact addressing.
Because of the inherent complexities involved, we believe actual traces of representative network traffic to be crucial in exploring the viability of network flow table cache management.

\subsubsection{Cache Simulation}
The simulator models a single inline device maintaining a stateful flow table, such as a firewall.
Maintaining this stateful flow table is the critical path for achievable throughput.
Caching techniques are leveraged to improve device throughput; this study focuses on improving the approach of cache management by borrowing techniques developed within the computer microarchitecture community.

\begin{table}[ht]
\small
\centering
\singlespacing
\begin{tabular}{|r|l|}
\hline
\textbf{Flow Cache Entries} & 
    4k flow entries \\ \hline
\textbf{Cache Associativity} &
    8-way \\ \hline
\textbf{Feature Tables} & \makecell[l]{
    5 features, each 64k entries \\
    5-bit saturating counters} \\ \hline
\textbf{Feedback History} & 
    8 predictions / associative-set \\ \hline
\textbf{Feature Selection Dataset} & 
    CAIDA Equinix Sanjose January 2012 \\ \hline
\textbf{Validation Datasets} & \makecell[l]{
    CAIDA Equinix Sanjose March 2012 \\
    CAIDA Equinix Chicago March 2014 \\
    CAIDA Equinix Chicago March 2016 \\
    CAIDA Equinix NYC March 2018} \\ \hline
\end{tabular}
\caption{Hashed Perceptron Cache Simulation Parameters}
\label{tbl:hp-cache-simulation}
\end{table}

Table \ref{tbl:hp-cache-simulation} summarizes the cache simulation parameters as well as feature selection and validation datasets.
The simulator replays PCAP files, preserving relative arrival order to maintain packet inter-arrival patterns.
The packet headers are parsed, stateful flow identifiers are maintained throughout the simulation to track and evaluate the cache management algorithm's relative performance.
In addition to cache hit-rates, the simulator gathers evaluation metrics helpful to gain insight during early design exploration of the \emph{Flow Correlator} cache management mechanism described in Section \ref{sec:hp-hp-design}.


We first provide insight into our feature exploration, evaluating several competing feature ranking mechanisms in Section \ref{sec:hp-exploration}.
Second, we evaluate the robustness of the final selected features across datasets in Section \ref{sec:hp-performance}.
In Section \ref{sec:hp-efficiency} we infer changes in cache entry lifetime by analyzing cache efficiency.
Finally, Section \ref{sec:hp-roles} recaps some of the dynamics between the chosen features in the context of bypass and reuse cache management predictions.

\subsection{Feature Exploration} \label{sec:hp-exploration}
While performance estimation is the ultimate goal of any architectural design-space exploration, quickly ranking useful features early in the design process is immensely valuable.
Feature design, ranking, and selection are time-consuming components of design exploration -- adding yet another dimension to an already complex optimization problem.

Since nearly all of the features explored in Table \ref{tbl:hp-all-features} perform poorly in isolation, it is crucial to compare feature performance as a system.
Further, it becomes non-trivial to select the best combinations of feature components as candidate combinations inherently share mutual information.
Therefore, there is a need for a reliable ranking mechanism that allows the architect to identify top-performing feature combinations early in the design phase.

The two competing feature ranking techniques analyzed in this section are Matthews Correlation Coefficient (MCC) and Total Influence (INF).
Randomly ordered features (RND) were also included to gauge the effectiveness of the mechanisms in context.
Finally, we describe a differential analysis technique developed in this work to trim mutual information, resulting in the final feature ranking.




\subsubsection{Feature Design} \label{sec:hp-feature-design}
\emph{Influence} was discovered out of a desire for feedback during feature design.
In particular, hints at the strengths and weaknesses of competing features help identify successful mixing strategies early in the design exploration.
The intuition behind Influence centers around interpreting the feature's correlation output (weight) as confidence.
Influence then contrasts these contribution confidences against the overall correctness of the system, grouped by prediction type (Reuse or Bypass).

\begin{figure}[t]
\centering
\includegraphics[width=0.9\linewidth]{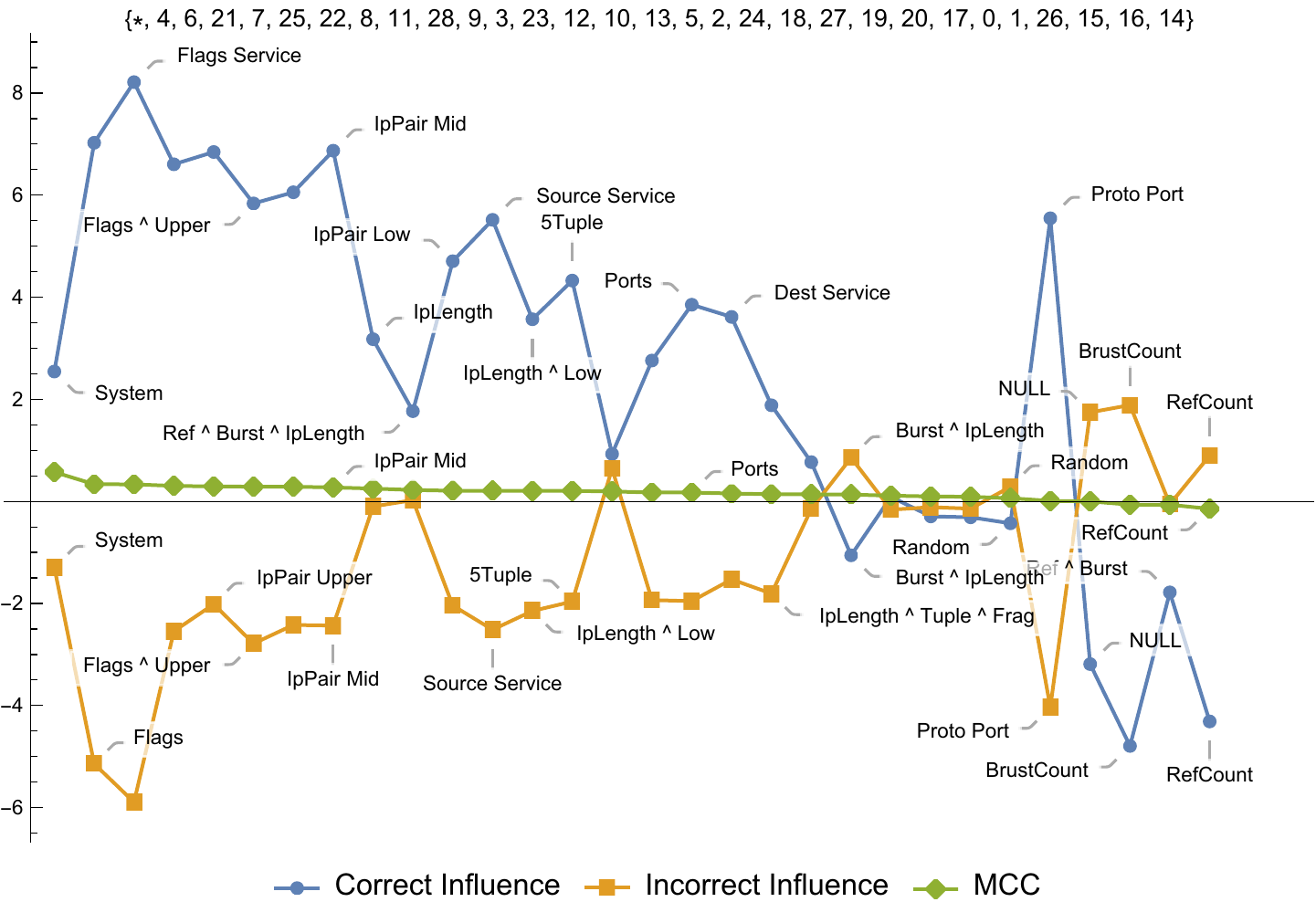}
\caption{Feature Influence on Reuse Predictions}
\label{fig:hp-reuse-mcc}
\end{figure}

Figure \ref{fig:hp-reuse-mcc} provides insight into feature contributions towards flow entry reuse predictions relative to MCC's balanced prediction accuracy.
Per-feature MCC and Influence metrics are tracked across all candidates in a single Hashed Perceptron \emph{Flow Correlation} simulation.
Features are sorted in decreasing MCC order with the corresponding \emph{Correct Influence} and \emph{Incorrect Influence} presented for relative comparison.

The first entry, \emph{System}, represents the overall flow entry reuse prediction exhibiting a notable improvement in MCC compared to any individual feature.
The contributions towards correct predictions (\emph{Correct Influence}) as well as contributions towards incorrect predictions (\emph{Incorrect Influence}) represent a feature's influence over the predictor.
The magnitude represents confidence, while the sign represents a positive (or negative) influence on predictor outcomes.

Interestingly, features exhibiting high confidence towards correct predictions also tend to be accompanied by overconfidence toward incorrect predictions.
Overconfidence can be tolerated when coupled with a high prediction accuracy (MCC); however, there is a strong preference for features that provide value while also minimally contributing to mispredictions as represented by \emph{Incorrect Influence}.


\begin{figure}[t]
\centering
\includegraphics[width=0.9\linewidth]{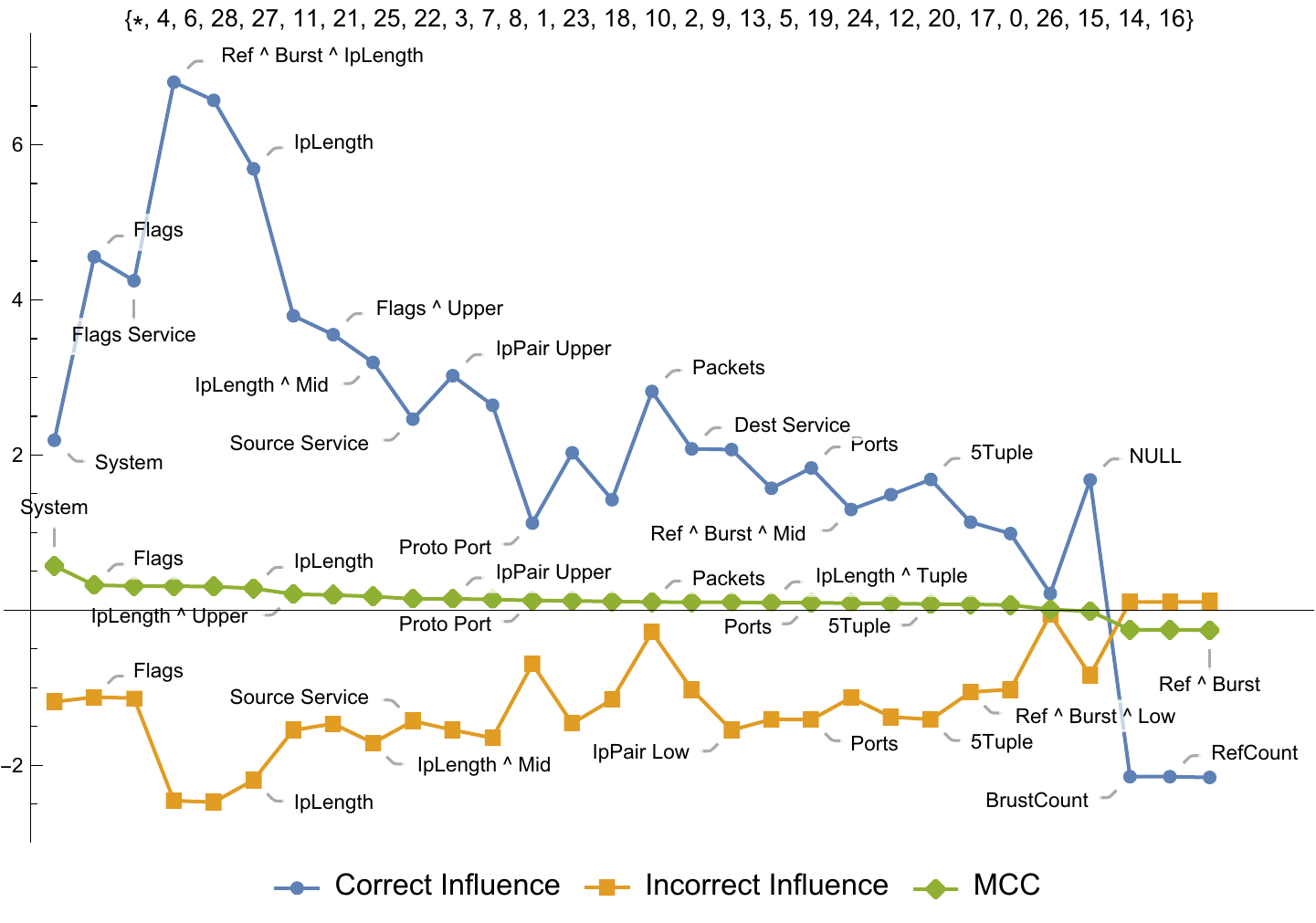} 
\caption{Feature Influence on Bypass Predictions}
\label{fig:hp-bypass-mcc}
\end{figure}

Figure \ref{fig:hp-bypass-mcc} provides insight into feature contributions towards flow entry bypass predictions relative to MCC.
It is intuitive that features have strengths towards either Bypass or Reuse predictions.
Unlike reuse flow entry predictions, bypass predictions inherently lack some of the temporal indicators tracked in cached flow table entries.

Bypass and Reuse predictions both aim to predict flow entry reuse within the approximation of the useful cache working set.
However, they are distinct predictions as the reuse probability of a cached and uncached flow table entry are not necessarily equivalent.
Bypass predictions occur not only at the start of new flows, but also on transitions from \emph{Dormant} to \emph{Active} for already established flows.
Reuse predictions aim to identify \emph{Active} connections amongst established flows, triggering early eviction if predicted to transition from \emph{Active} to \emph{Dormant}.

Burst Count and Reference Count (Pure features $f_{14}$ and $f_{15}$) exhibit a perplexing anti-correlation as seen by both MCC and Influence.
It is notable that Burst and Reference Count rely on unrolled and allowed to decouple from global patterns.
Intuitively, it makes sense that temporal patterns need to be grouped (unrolled) in some way to enable correlations.
During feature design, Influence enabled insight into how each feature fell into natural cache management roles.




\subsubsection{Feature Ranking} \label{sec:hp-ranking}

\begin{figure}[ht]
\centering
\includegraphics[width=\linewidth]{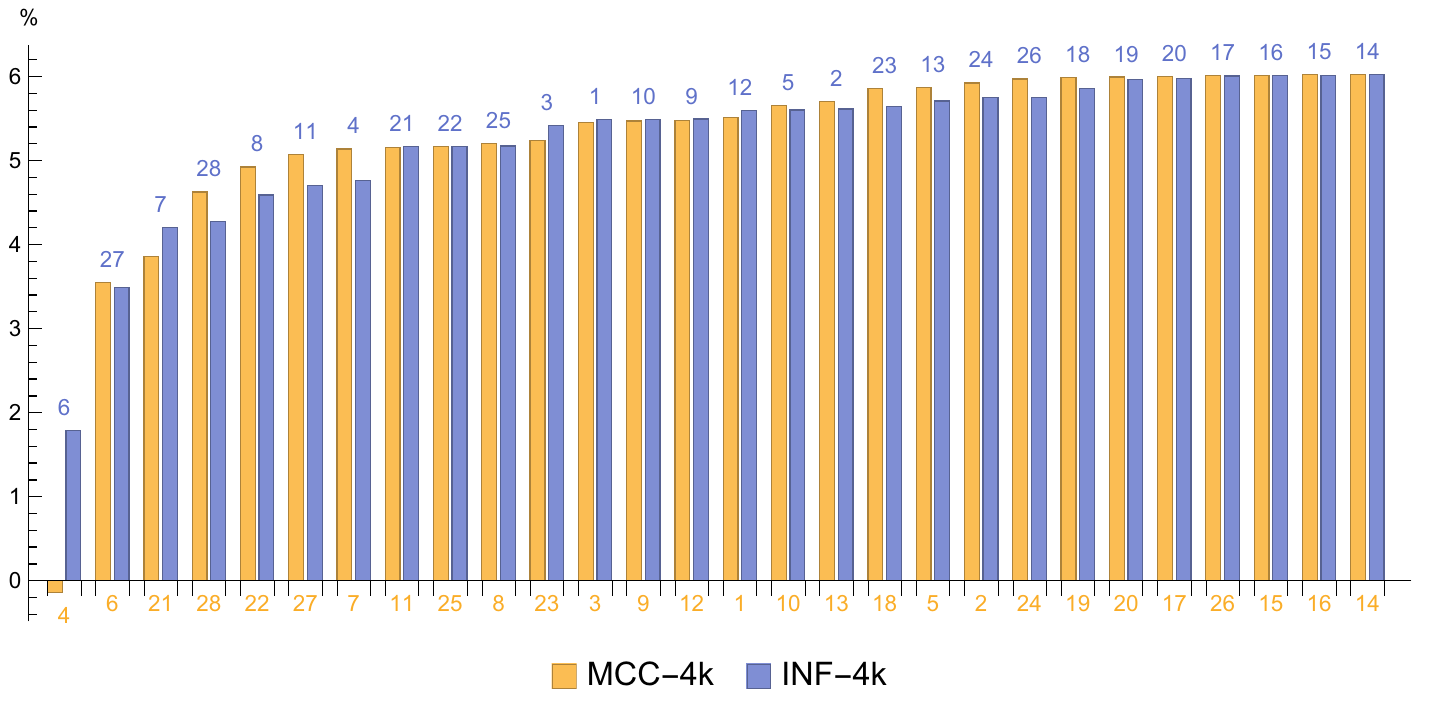} 
\longcaption{Initial Feature Ranking}{
Hit-rate improvement over baseline LRU comparing MCC-4k below bars in yellow to INF-4k above in blue.}
\label{fig:hp-rank-initial}
\end{figure}

Figure \ref{fig:hp-rank-initial} shows the relative hit-rate improvement over LRU as baseline.
\emph{Total Influence} as defined in \ref{eq:f-total-inf} was used for INF-4k's ranking.
The plot is generated by incrementally including features in ranked order using Algorithm \ref{alg:hp-sweep}.
This \emph{sweep} plot shows the incremental gain in hit rate by including additional features as a useful means to compare feature rankings.
Spanning from a single feature (left) to all features (right), diminishing returns can be easily identified.





Both MCC and INF provide a relatively similar ranking, with a few notable differences.
In particular, MCC's top choice, $f_{4}$ consists of just TCP flags is unable to stand alone.
INF's top choice, $f_{6}$, consists of a mix of Port, Protocol, and TCP/IP flags, covering $f_{4}$.
As a single-table predictor, INF's $f_{6}$ manages to provide a modest 2\% hit-rate improvement over baseline LRU, while MCC's $f_{4}$ actually degrades hit-rate.
Ultimately MCC provides a better all-rounded ranking; however, INF does identify a few notably valuable features before MCC.

\begin{figure}[H]
\centering
\includegraphics[width=\linewidth]{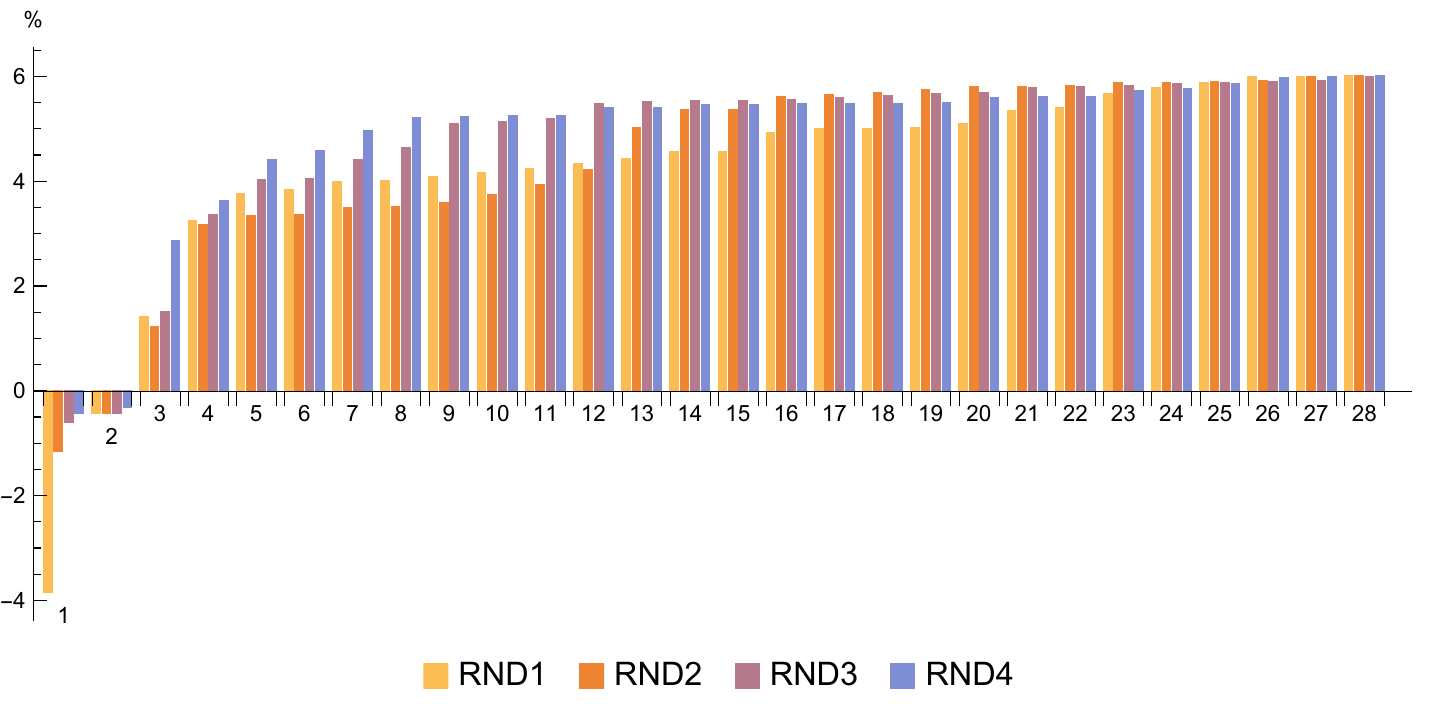}
\caption{Random Feature Ranking}
\label{fig:hp-rank-random}
\end{figure}

To help put into perspective the utility of MCC as a feature ranking mechanism, we compare it against four randomized rankings.
As expected, Figure \ref{fig:hp-rank-random} showcases the stunted performance of random rankings amongst features with significant permutation overlap.
There are a few notable observations that highlight the behavior of a Hashed Perceptron-based predictor.

The first notable observation is that predictor effectiveness under-performs baseline LRU until the third randomly selected feature is added.
The situation flips with the third random selection able to surpass the baseline; achieving half the potential hit-rate improvement by the fourth random selection.

As expected, there is significant shared information across the features in Table \ref{tbl:hp-all-features} due to feature permutation similarity.
The goal of the ranking algorithm is to identify the best combinations in order to hasten designer iteration.
It is clear that MCC and INF are both able to identify a better initial ranking than random selections.
However, MCC and INF are both unable to account for shared information, looking only at the feature's own track record in isolation.

\subsubsection{Iterative Information Gain} \label{sec:hp-rank-ig}
Relying on the designer to select the best choice is untenable as system complexity grows.
Ultimately we need a method to rank features by incremental value contributed to the system.
Inspired by information gain decision tree learning algorithms, we outline our approach to both measure a feature's actual contribution to the system as well as take into account mutual information between features.

\begin{figure}[H]
\centering
\includegraphics[width=\linewidth]{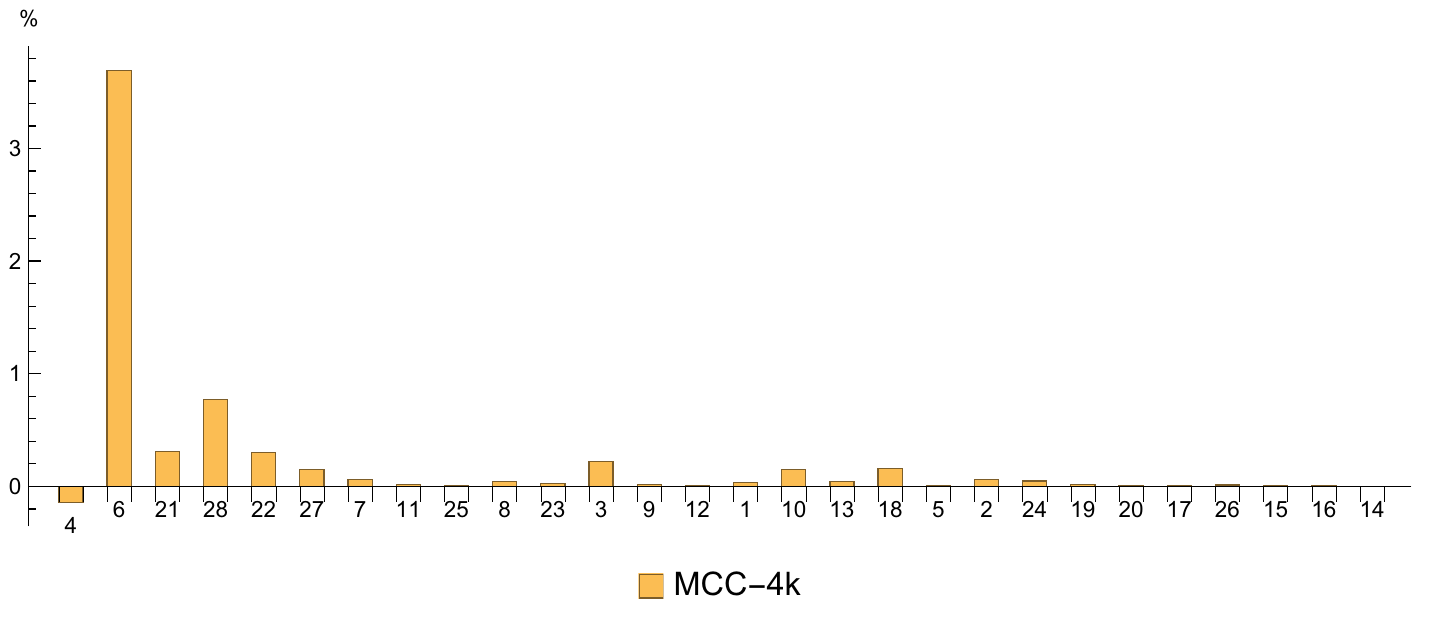}
\longcaption{Differential Improvement}{
Per-feature hit-rate differential gain (adjacent difference) of MCC-4k.}
\label{fig:hp-rank-differential}
\end{figure}

Our iterative information gain method outlined in Section \ref{sec:hp-information-gain} improves an MCC ranking by differential analysis sweep.
The feature sweep analysis plot is differential analysis, where hit rate is plotted as features are incrementally added until all features are included.
Figure \ref{fig:hp-rank-differential} shows the incremental change in hit rate as features are incrementally added to the predictor in MCC-4k rank order.

Figure \ref{fig:hp-rank-differential} shows the change in hit-rate as features are added to the predictor from left to right.
While early features are likely to bring a larger gain, the goal is to minimize shared information between features.

\begin{figure}[H]
\centering
\includegraphics[width=\linewidth]{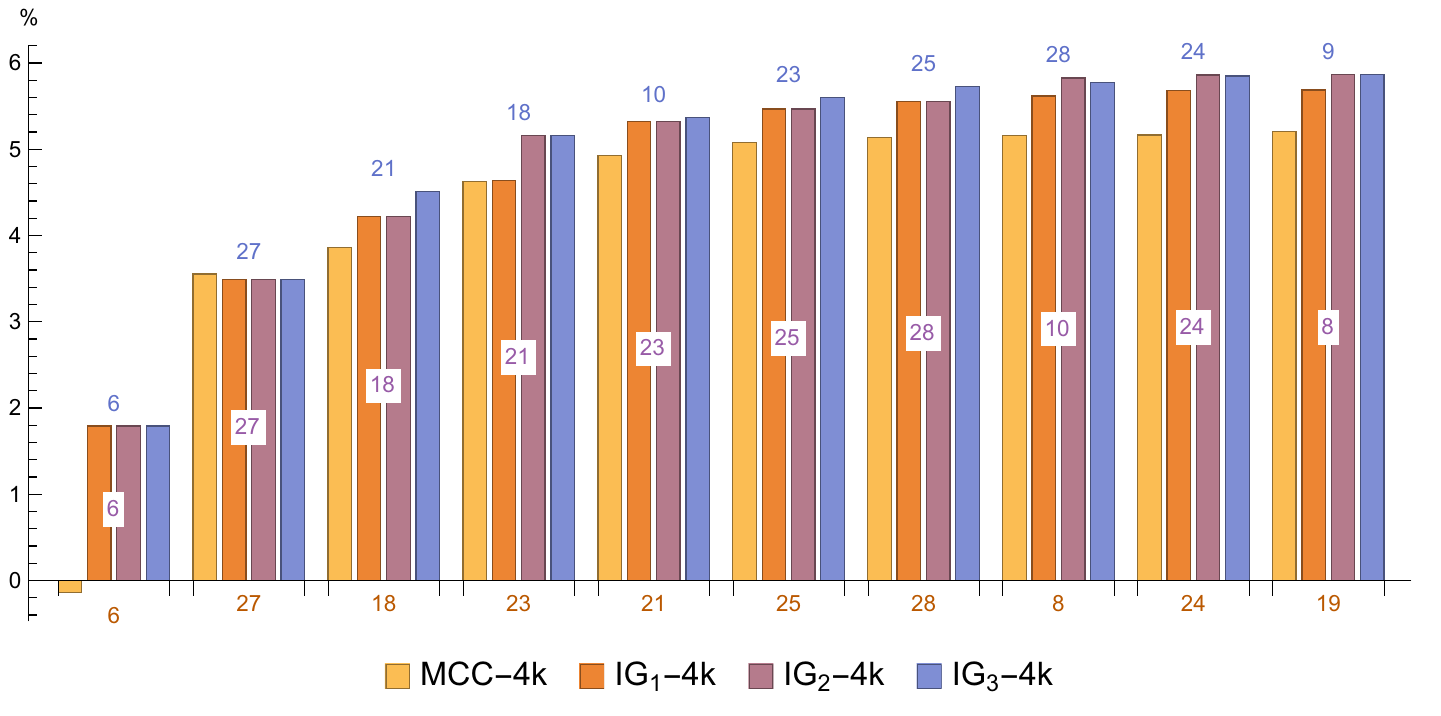} 
\caption{Iterative Information Gain Ranking Improvement}
\label{fig:hp-rank-ig}
\end{figure}

Figure \ref{fig:hp-rank-ig} shows the iterative improvement from the initial \emph{MCC-4k} ranking through three iterations of differential information gain.
To improve readability, this figure just showcases the hit-rate improvement for the first ten features.
\emph{IG\textsubscript{1}-4k} through \emph{IG\textsubscript{3}-4k} represent the improvement in ranking through each differential Information Gain (IG) iteration.

The first notable adjustment made in the first pass (IG\textsubscript{1}) is identifying that $f_{6}$ covers $f_{4}$ entirely and translates to a hit-rate improvement as the first feature, despite $f_{4}$ achieving a marginally higher prediction accuracy as inferred by MCC.

It is also notable that $f_{10}$ and $f_{18}$ were both overlooked by MCC and INF, implying a relatively low prediction accuracy.
However, both propagated to the top five ranking in the subsequent two iterations (IG\textsubscript{2} and IG\textsubscript{3}).
This differential information gain analysis identified that $f_{10}$ and $f_{18}$ provided notable value to the system, despite the lower prediction accuracy.

The final notable adjustments were marginal, preferring to unroll across upper IP addresses rather than middle or lower variants ($f_{18}$ and $f_{21}$).

This iterative information gain technique is best fit to trim shared information from an already decent ranking.
While not shown here, attempts to iterate from a random feature selection using just the iterative information gain technique were computationally prohibitive.

We found Influence to be particularly useful during feature creation, MCC best suited for initial feature ranking, and differential information gain notably helpful to narrow in on the most informative features.
Automating feature selection, even if imperfect, allows the system designer to focus on feature creation -- in particular, introducing novel feature combinations and information sources.

\subsection{Improvement Validation} \label{sec:hp-performance}



\begin{figure}[ht]
\centering
\includegraphics[width=\linewidth]{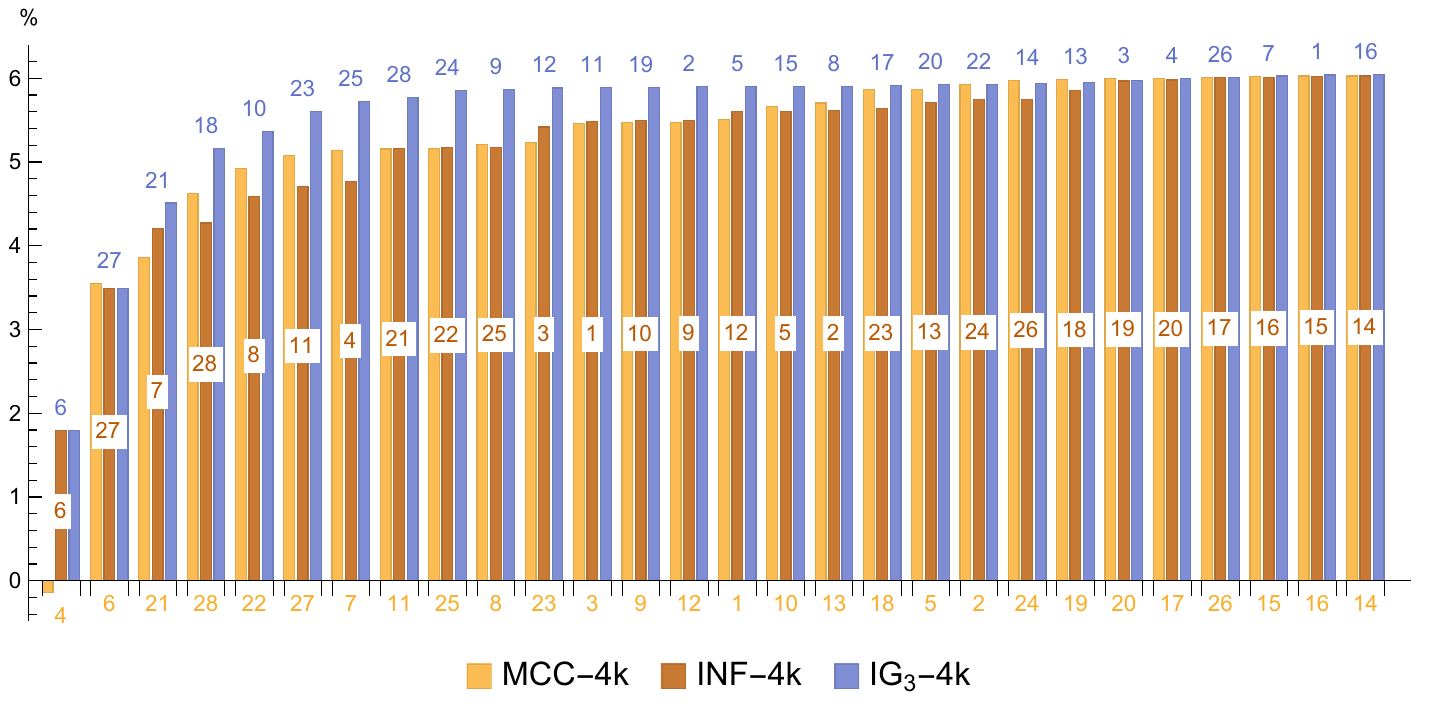} 
\caption{Final Feature Ranking Comparison}
\label{fig:hp-rank}
\end{figure}

Figure \ref{fig:hp-rank} recaps the ranking improvement on the training dataset.
Diminishing returns in achievable hit-rate improvements started to occur roughly between four and seven features.
For validation, we decided to select the five best features from each ranking.


\begin{table}[H]
\vspace{-\baselineskip}
\small
\centering
\singlespacing
\begin{tabular}{cll}
\textbf{$f_{\#}$} & \textbf{Feature Type} & \textbf{Feature Components} \\
6  & Pure Mix & $f_{TCP}\!\ll\!7 \wedge ipProto \wedge min(srcPort, dstPort)$ \\
27 & Pure + Pattern & $\{ipLength, \lceil burstCount \rceil_8 \}$ \\
21 & Pure + Pattern & $ipLength \wedge (ipv4Dst \wedge ipv4Src)[31\!\colon\!16]$ \\
18 & Pure + Pattern & $\{\lceil refCount \rceil_8, \lceil burstCount \rceil_8\} \wedge (ipv4Dst \wedge ipv4Src)[31\!\colon\!16]$ \\
10 & Pattern & $\lceil flowPackets \rceil_{16}$
\end{tabular}
\longcaption{Selected Features (IG\textsubscript{3}-4k)}{
Optimized on flow cache with 4k entries, 8-way set associativity using the CAIDA Equinix Sanjose January 2012 dataset (equinix-sanjose.20120119) during feature selection and ranking.}
\label{tbl:hp-selected-features}
\end{table}

Table \ref{tbl:hp-selected-features} list the top five features in ranked order of improved system value after the third iteration of differential information gain (IG\textsubscript{3}-4k).
It is interesting to note that all \emph{Pure} information sources were utilized in some form in the final set of features.
While it is not surprising that protocol and TCP/IP flags proved to be naturally useful reuse hints, it is certainly interesting that $f_{6}$ was preferred over the $f_{4}$.
A similarly useful reuse hint comes from deviations in IP frame length.
While both MCC and INF showed high utility of frame length ($f_{11}$) as a \emph{Pure} indicator, IG\textsubscript{3} found significantly higher utility by combining frame length and BurstCount ($f_{27}$).

It appears that both frame length and BurstCount contributed uniquely as a reuse prediction hint, both notably providing additional value being unrolling over upper IP addresses ($f_{18}$ and $f_{21}$).
With IP address combinations seemingly contributing little insight into reuse as \emph{Pure} sources, it is interesting that both frame length and BurstCount found extra utility through unrolling.






\subsubsection{Validation Datasets}
While it is clear that there is notable improvement performing IG passes within the training dataset, it is crucial that this process did not over-optimize feature selection.
The training set consisted of five minutes of PCAP, bi-directional from January 2012, taken at CAIDA Sanjose exchange.
The validation set consisted of five minutes of PCAP, bi-directional from March of 2012, 2014, 2016, and 2018, spanning the three available CAIDA exchange locations: Sanjose, Chicago, and NYC.
The validation set will be used to compare the top-five features provided by the MCC-4k, INF-4k, and IG\textsubscript{3}-4k rankings.

\begin{figure}[H]
\centering
\includegraphics[width=0.7\linewidth]{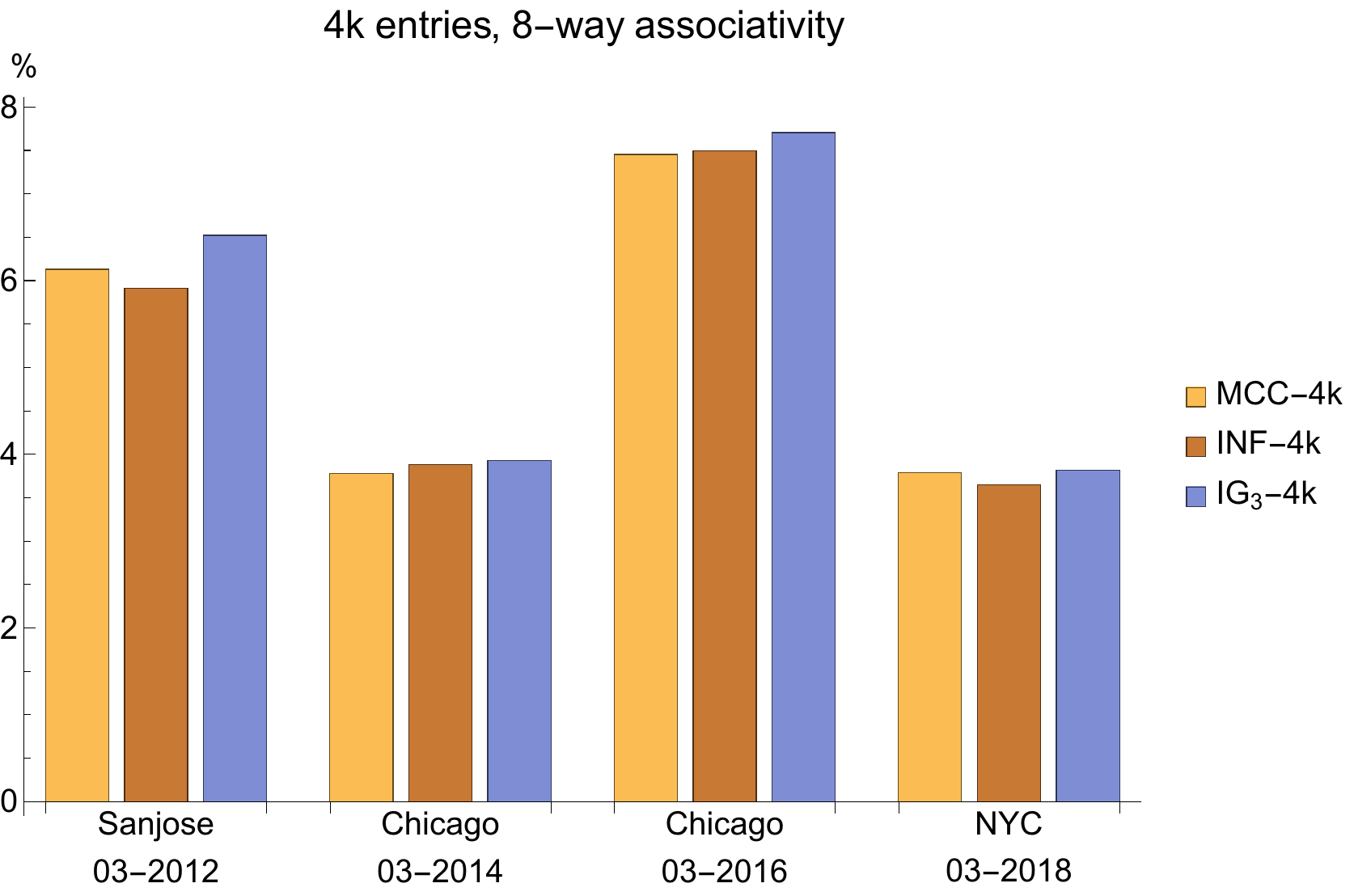}
\caption{Improved Hit-Rate Validation}
\label{fig:hp-validation}
\end{figure}

Figure \ref{fig:hp-validation} shows consistent improvement across all validation sets.
Hit-rate improvements ranged between 4\% and just under 8\% over baseline LRU.
The validation datasets showed no degradation from MCC-4k to IG\textsubscript{3}-4k, indicating that iterative Information Gain did not over-optimize.
However, we suspect IG's greedy optimization requires generality in feature creation -- best suited to remove mutual information rather than select the most robust features.


These results attest to the generality of the features chosen, providing value to a generic cache management reuse predictor beyond the original dataset location.
It is certainly notable for demonstrating viability towards a flow table cache management approach that translates beyond the original optimization parameters.






\subsection{Cache Efficiency} \label{sec:hp-efficiency}

\begin{figure}[H]
\centering
\includegraphics[width=55mm]{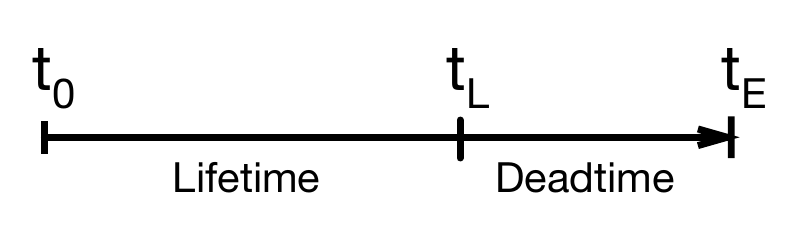}
\longcaption{Cache Entry Lifecycle}{
$t_0$: insertion time, $t_L$: last-access time, $t_E$: eviction time}
\label{fig:hp-entry-life}
\end{figure}

Tracking the lifecycle of cache entries can be helpful to understand how cache management techniques impact cache behavior.
Figure \ref{fig:hp-entry-life} depicts a cache entry's lifecycle from insertion ($t_0$) to last-access time ($t_L$) and finally the eviction time ($t_E$).

Equation \ref{eq:f-lt} defines a cache line's \emph{lifetime} as duration between insertion and last access.

\begin{equation}
lifetime = t_L - t_0
\label{eq:f-lt}
\end{equation}

Similarly, \emph{deadtime} is the duration between last access and eviction as defined in Equation \ref{eq:f-dt}.

\begin{equation}
deadtime = t_L - t_E
\label{eq:f-dt}
\end{equation}

Finally, Equation \ref{eq:f-efficiency} defines cache line \emph{efficiency} as the useful \emph{lifetime} over the total time the cache entry was occupied.

\begin{equation}
efficiency = \frac{t_L - t_0}{t_E - t_0}
\label{eq:f-efficiency}
\end{equation}

Tracking the \emph{lifetime}, \emph{deadtime}, and \emph{efficiency} for all cache lines throughout the simulation provides some insight into the relative impact of cache management algorithms on overall caching behavior.

\begin{figure}[htb]
\centering
\includegraphics[width=0.8\linewidth]{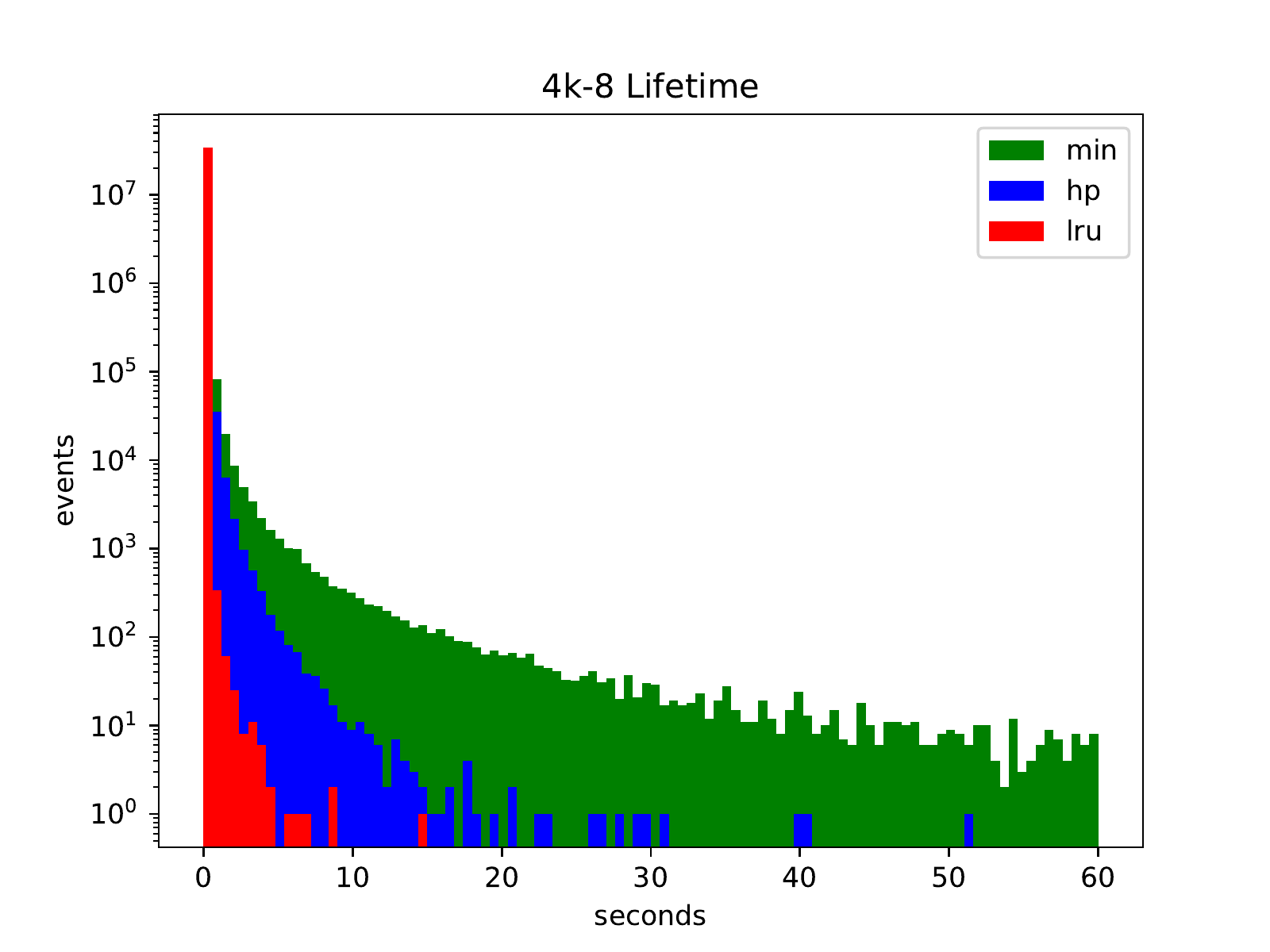} 
\caption{Cache Entry Lifetime}
\label{fig:hp-lifetime}
\end{figure}

Figure \ref{fig:hp-lifetime} compares the cache line \emph{lifetime} during a cache simulation over a one-minute PCAP trace\footnote[1]{CAIDA Equinix Sanjose January 2012 PCAP trace: equinix-sanjose.20120119.}.
While the vast majority of flow entries have a short lifetime in the cache, the \emph{Flow Correlator} Hashed Perceptron cache management technique (hp) more closely resembles Belady's MIN (min).
Holding onto certain bursty flows for a longer duration between pauses potentially reduces entry turn-over.

\begin{figure}[htb]
\centering
\includegraphics[width=0.8\linewidth]{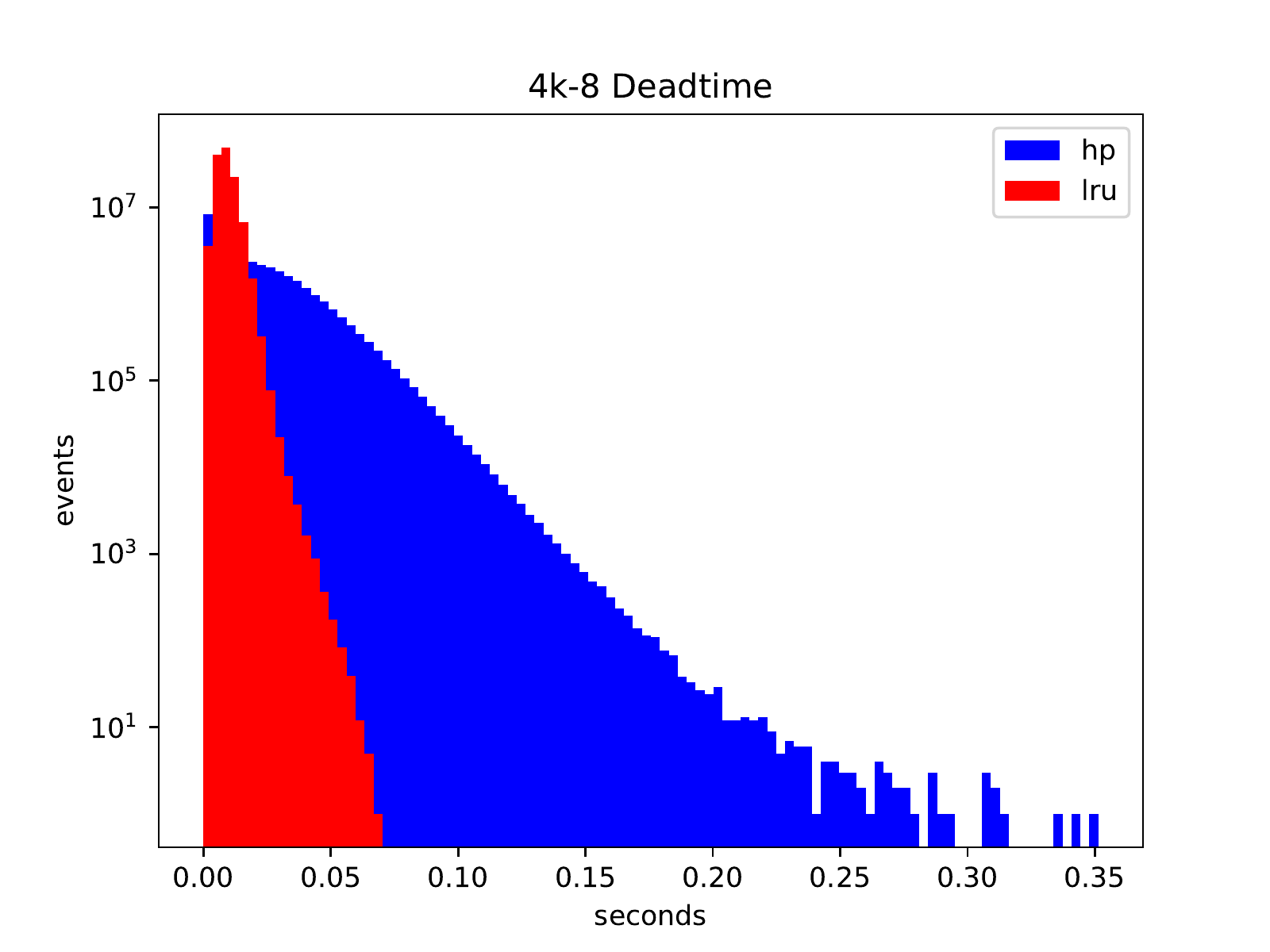} 
\caption{Cache Entry Deadtime}
\label{fig:hp-deadtime}
\end{figure}

Figure \ref{fig:hp-deadtime} shows the corresponding \emph{deadtime}, implying that the hashed perceptron technique simultaneously has both more and less patience.
A significant number of entries are evicted very early (within 10ms), while simultaneously waiting on average twice as long as LRU for others (upwards of 200ms).
This adaptability allows hp to triage flows with a low probability of reuse, while simultaneously being willing to increase entry \emph{deadtime} in hopes of also enabling an increase in entry \emph{lifetime}.

\begin{figure}[htb]
\centering
\includegraphics[width=0.8\linewidth]{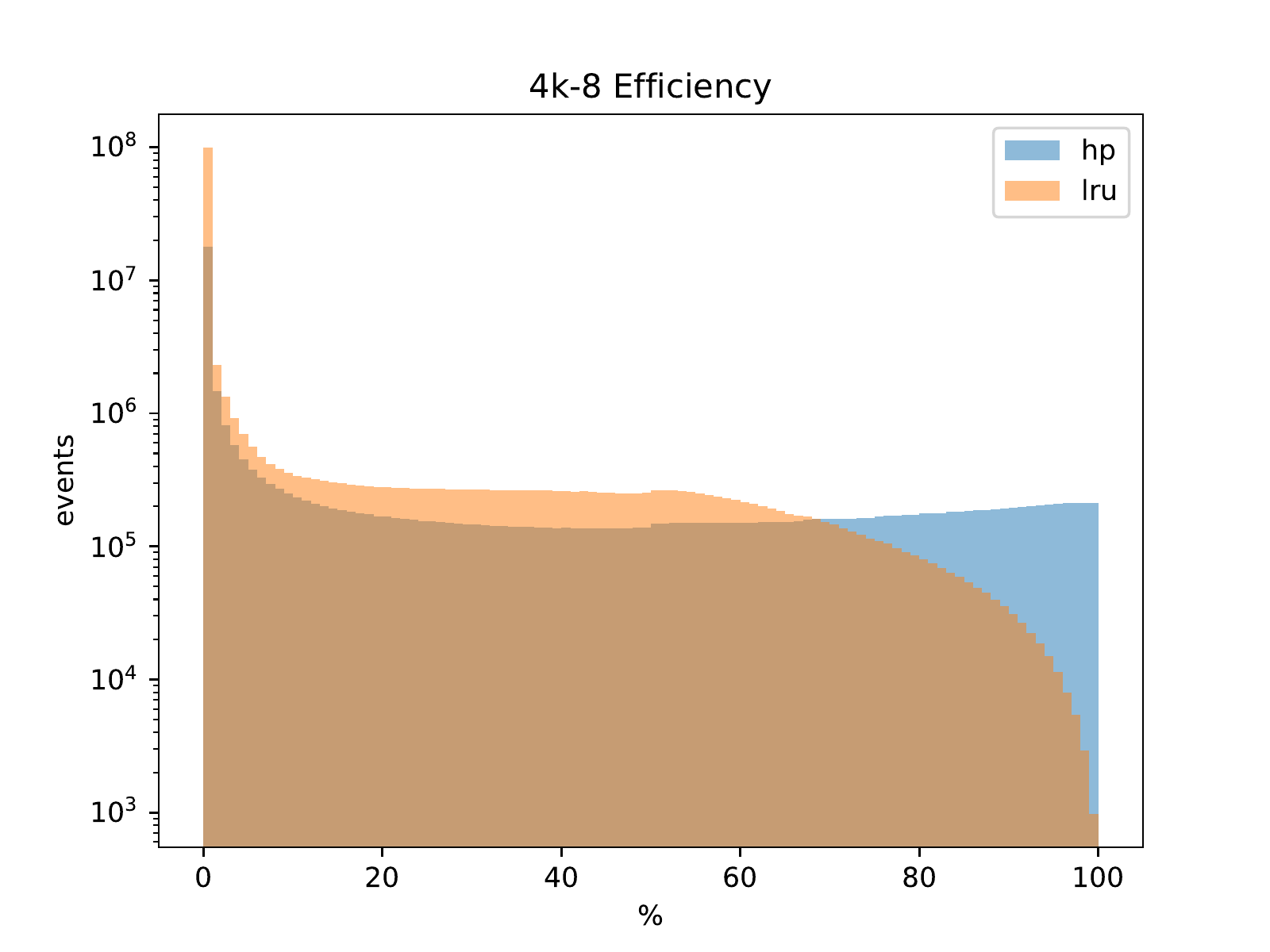}
\caption{Cache Entry Efficiency}
\label{fig:hp-efficiency}
\end{figure}

Figure \ref{fig:hp-efficiency} outlines how the hashed perceptron approach generally smoothed out cache \emph{efficiency}.
LRU's inherent requirement to wait for demotion from MRU to LRU limits cache efficiency.
Meanwhile, the hashed perceptron technique is a step towards both triaging flows with a low probability of reuse as well as a tendency to hold onto cache entries for a longer duration.

Cache \emph{efficiency} doesn't grant perspective into the number of hits over the entry \emph{lifetime}, just that the entry had reuse.
Coupled with an improvement to hit rate, this cache line efficiency analysis confirms that there is a worthwhile trade-off to holding onto certain flows longer as well as triaging flows early.


\subsection{Feature Roles} \label{sec:hp-roles}
Throughout the feature selection process, we noticed a few noteworthy trends in roles assumed by each feature.
Observed primarily through feature influence described in Section \ref{sec:hp-influence}, features tended to produce a certain prediction bias as measured by average output prediction weight.
Originally described as a form of optimism, features tended to vote for eviction (pessimistic), while others tended to lean towards reuse (optimistic).
What was observed as a natural feature bias towards a particular outcome in actually reflects the natural predictive utility of the feature.

Certain features are better indicators of early eviction -- hinting at a flow's transition from \emph{active} to \emph{dormant}.
The temporal pattern feature components ($f_{10}$, $f_{14}$, and $f_{15}$) are clear contributes to bypass role.
Other features are best suited to triage incoming flows during bypass -- hinting at flow transitions from \emph{dormant} to \emph{active}.
Indications of protocol behavior from several pure feature components ($f_{4}$, $f_{6}$, and $f_{11}$) were notable contributions to the reuse role.

These two cache management roles (bypass and reuse) tend to uniquely favor subsets of feature components.
In nearly all cases, we noticed an overall benefit in combining feature components with orthogonal roles.
Additionally, we also noticed that certain orthogonal combinations ($f_{27}$, $f_{28}$) resulted in improved prediction accuracy and confidence for both bypass and reuse.
IP frame length ($f_{11}$), while valuable as a pure feature component contributing to both roles, strengthened nearly any feature component it was combined with.

As an artifact of packetizing flows into manageable units bounded by Maximum Transmission Unit (MTU), IP's frame length field was a notable indicator of packet inter-arrival patterns.
It is further noteworthy that while $f_{11}$ was able to stand alone, it also significantly strengthened when combined with role-focused feature components.
While not included in this study, the interplay between frame length and received packet size (only differing if IP fragmentation has occurred), is expected to be noteworthy low-hanging fruit in further feature design explorations.
The final feature ranking pass of our differential Information Gain approach preferred the information density of role-orthogonal feature components.


\subsection{Automatic Throttling}
One of the critical aspects of cache management is the ability to dynamically adapt prediction aggressiveness in response to changing cache pressure.
While not quite hardened against adversarial attacks, the Hash Perceptron flow correlation mechanism presented in this work exhibits inherent self-throttling behavior which adapts to changes in cache pressure.
The correlation feedback mechanism reinforces likely good predictions, while predictions with less probability of being correct naturally tend to regress to the natural bias of the system.  

\begin{figure}[ht] \centering
\includegraphics[width=0.9\linewidth]{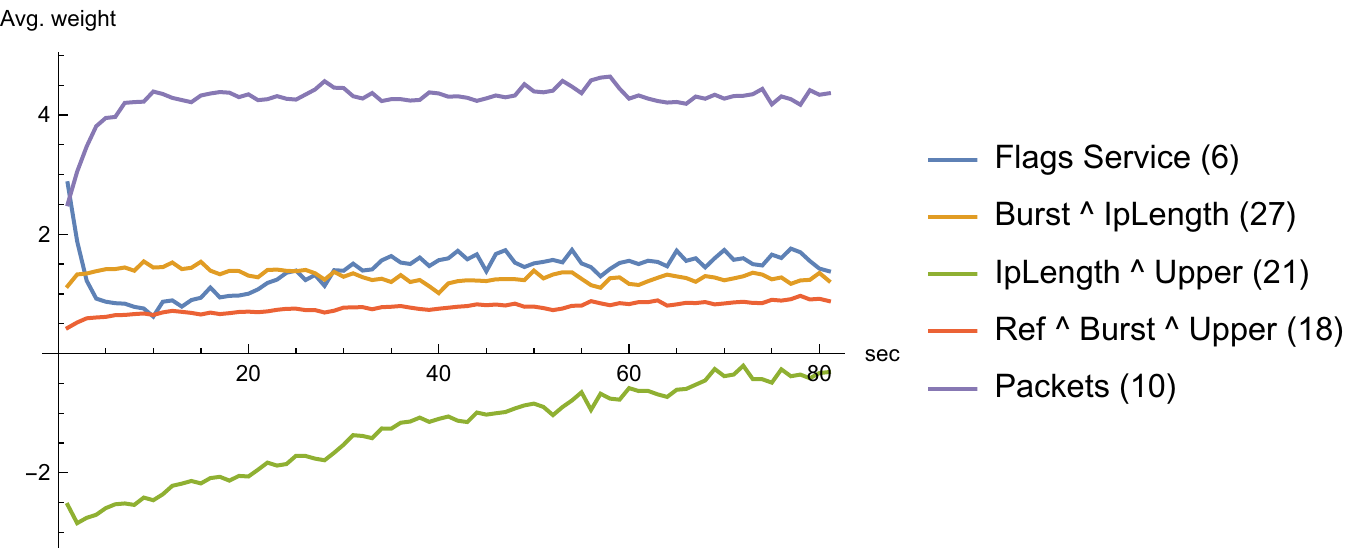}
\caption{Feature Bias}
\label{fig:hp-rank-bias}
\end{figure}

Figure \ref{fig:hp-rank-bias} shows the feature's prediction bias over epochs of each evaluated feature.
Naively, we would desire all features to be neutral-biased; however, this isn't aligned with the objective of discriminating flows based on estimated short-term activity.
Features assume a role implicit by the construction of their components, producing a bias skewed towards the utility of their insight.
Pessimistic features (net bias of evict) primarily contribute to the role of bypass, effectively filtering flows from entering the cache until sufficient optimism arises from other features.
Optimistic features (net bias of keep) primarily contribute to reuses prediction.
The adaptive nature of hashed perception mechanism dynamically adjusts each feature to the average observed cache pressure.

\section{Summary}
Cache management for stateful flow tables is a largely unexplored area of research, despite their wide deployment.
This work demonstrates the viability of the Hashed Perceptron approach to cache management.
The Hashed Perceptron correlation technique is uniquely able to cope with noisy information sources, well suited for network flow table cache management.


Feature creation is still a designer's art coupled with both intuitive and counter-intuitive surprises.
Early feedback and iteration are essential to enable fast design exploration.
Advancements to systematic feature ranking and selection, even if non-optimal, allow designers to focus on system architecture and feature creation first and foremost.
Ultimately, any ML approach is inherently reliant on effective feature creation and information sources.

The Hashed Perceptron technique is proving to be an interesting lightweight ML approach where weights are run-time correlations, enabling adaptability beyond comparable offline techniques.
When applying the Hashed Perceptron technique to a new domain it became apparent how a multi-feature, consensus-based system significantly improves prediction reliability compared to hand-crafting complex heuristics.
An interesting observation in the context of network traffic is the resiliency of the Hashed Perceptron technique to cope with noisy or even misleading features.
While it is certainly possible to degrade performance, the Hashed Perceptron's ability to combine multiple features offers stability over any single perspective technique.


Few mechanisms allow combining independent correlations reliably, yet efficiently to improve system performance.
This work describes the process of adapting and evaluating the Hashed Perceptron mechanism to an unexplored domain, where the input patterns and vectors available to features bring unique challenges.

\bibliographystyle{ieeetr}
\bibliography{ref_networking, ref_caching}

\end{document}